\documentclass[aps,pra,
   reprint,
   superscriptaddress,
   showpacs,
   preprintnumbers,
   ]{revtex4-1}
\usepackage[T1]{fontenc}          % this package is necessary for T1 extended fonts (\dj)
\usepackage{amsfonts}
\usepackage{amssymb}
\usepackage{amsmath}
\usepackage{color}
\usepackage{graphicx}
\usepackage{epsfig}
\usepackage{subfigure}
\usepackage{dcolumn}
\usepackage{time}
\usepackage[pdftex,colorlinks=true,linkcolor=red,citecolor=blue]{hyperref}
\definecolor{dark}{gray}{.5}
\definecolor{light}{gray}{.75}
\definecolor{darkmagenta}{rgb}{.5,0,.5}
\newcommand{\Veff}{\mathcal{V}_{\text{eff}}}             % effective potential
\newcommand{\Veffzero}{\mathcal{V}_0}                    % zero effective potential
\newcommand{\rD}{\text{D}}

\newcommand{\bk}{\mathbf{k}}

\newcommand{\br}{\mathbf{r}}

\newcommand{\calD}{\mathcal{D}}

\newcommand{\calG}{\mathcal{G}}

\newcommand{\calL}{\mathcal{L}}

\newcommand{\calN}{\mathcal{N}}
\newcommand{\calO}{\mathcal{O}}

\newcommand{\calS}{\mathcal{S}}

\newcommand{\calV}{\mathcal{V}}

\newcommand{\hH}{\hat{H}}

\newcommand{\hN}{\hat{N}}

\newcommand{\kBolt}{k_{\text{B}}}              % Boltzmann's constant
\newcommand{\kF}{k_{\text{F}}}                 % Fermi momentum
\newcommand{\epsilonF}{\epsilon_{\text{F}}}    % Fermi energy
\newcommand{\vs}{vs.}                      % vs  
\newcommand{\QED}%
   {$\mathcal{Q\kern-.1em \lower.6ex\hbox{$\mathcal{E}$}\kern-.1667em D}$}
\newcommand{\QCD}%
   {$\mathcal{Q\kern-.1em \lower.6ex\hbox{$\mathcal{C}$}\kern-.1667em D}$}
\newcommand{\SUSYext}%
   {$\mathcal{S \kern-0.08em \lower 0.5ex \hbox{$\mathcal{U}$}
   \kern-0.05em S \kern-0.2em \lower 0.5ex
   \hbox{$\mathcal{Y}$}}\kern-0.05em{}_{\text{ext}}$}
\newcommand{\Quad}[1]{\quad\text{#1}\quad}         % small spacer
\newcommand{\Qquad}[1]{\qquad\text{#1}\qquad}      % big spacer
\newcolumntype{d}[1]{D{.}{.}{#1}}
\newcommand{\Set}[1]{ \bigl ( \, #1 \, \bigr )}    % Set
\newcommand{\Diag}[1]{\mathrm{diag}( \, #1 \, )}   % Diag
\newcommand{\rd}{\mathrm{d}}
\newcommand{\Partial}[4]%                          % partial derivatives
   {\Bigl ( \frac{\partial #1 }{\partial #2 } \Bigr )_{\! #3, #4 }}
\newcommand{\Intk}{\int% 
   \frac{ \mathrm{d}^3 k }{ (2\pi)^3 } }           % Int d^3 k / (2\pi)^3

\newcommand{\Det}[1]{\det [ \, #1 \, ]}

\newcommand{\Ln}[1]{\ln [ \, #1 \, ]}

\newcommand{\Sgn}[1]{\mathrm{sgn}[ \, #1 \, ]}     % sgn
\newcommand{\Tr}[1]{\mathrm{Tr} [ \, #1 \, ]}
\newcommand{\Trb}[1]{\mathrm{Tr} \bigl \lbrack \, #1 \, \bigr \rbrack }

\newcommand{\bra}[1]%
   {\ensuremath{\langle \, #1 \, |}}
\newcommand{\Bra}[1]%
   {\ensuremath{\langle \, #1 \, |}}
\newcommand{\bigbra}[1]%
   {\ensuremath{\Bigl \langle \, #1 \, \Bigr |}}
\newcommand{\ket}[1]%
   {\ensuremath{| \, #1 \, \rangle}}
\newcommand{\Ket}[1]%
   {\ensuremath{| \, #1 \, \rangle}}
\newcommand{\bigket}[1]%
   {\ensuremath{\Bigl | \, #1 \, \Bigr \rangle}}
\newcommand{\braket}[2]%
   {\ensuremath{\langle \, #1 \, | \, #2 \, \rangle}}
\newcommand{\Braket}[2]%
   {\ensuremath{\langle \, #1 \, | \, #2 \, \rangle}}
\newcommand{\matrixelement}[3]%
   {\ensuremath{\langle \, #1 \, | \, #2 \, | \, #3 \, \rangle}}
\newcommand{\MatEl}[3]%
   {\ensuremath{\langle \, #1 \, | \, #2 \, | \, #3 \, \rangle}}
\newcommand{\pbra}[1]%
   {\ensuremath{( \, #1 \, |}}
\newcommand{\pket}[1]%
   {\ensuremath{| \, #1 \, )}}    
\newcommand{\pbraket}[2]%
   {\ensuremath{( \, #1 \, | \, #2 \, )}}
\newcommand{\braV}[1]%
   {\ensuremath{\langle \, #1 \, \Vert}}
\newcommand{\ketV}[1]%
   {\ensuremath{\Vert \, #1 \, \rangle}}
\newcommand{\Comm}[2]%
   {\ensuremath{[ \, #1, #2 \, ]}}
\newcommand{\AntiComm}[2]%
   {\ensuremath{\{ \, #1, #2 \, \}}}
\newcommand{\Pbracket}[2]%
   {\ensuremath{\{ \, #1, #2 \, \} }}
\newcommand{\PBracket}[2]%
   { \ensuremath{ \{ \, #1, #2 \, \}_{\lower1.0ex\hbox{\scriptsize \text{PB}}} } }
\newcommand{\wedgeComm}[2]
   {\ensuremath{[ \, #1, #2 \, ]_{\lower1.0ex\hbox{\scriptsize $\wedge$}} }}
\newcommand{\Expect}[1]%                           % Expect
   {\ensuremath{\langle \, #1 \,  \rangle}}
\newcommand{\expect}[1]%
   {\ensuremath{\langle \, #1 \,  \rangle}}
\newcommand{\expectbig}[1]%
   {\ensuremath{\Bigl \langle \, #1 \, \Bigr \rangle}}
\newcommand{\expectc}[2]%
   {\ensuremath{\langle \, \{ \, #1 , #2 \, \} \, \rangle}}
\newcommand{\expectq}[2]%
   {\ensuremath{\langle \, [ \, #1 , #2 \, ] \, \rangle}}
\newcommand{\expectT}[1]%
   {\ensuremath{\langle \, \mathcal{T} \{ \, #1 \, \} \, \rangle}}
\newcommand{\expectaT}[1]%
   {\ensuremath{\langle \, \mathcal{T}^{\ast} \{ \, #1 \, \} \, \rangle}}
\newcommand{\expectTbig}[1]%
   {\ensuremath{\biggl \langle \, \mathcal{T}  \biggl \{ \, #1 \,%
\biggr \} \, \biggr \rangle}}
\newcommand{\expectTabig}[1]%
   {\ensuremath{\biggl \langle \, \mathcal{T}^{\ast} \biggl \{ \, #1 \,%
\biggr \} \, \biggr \rangle}}
\newcommand{\Tproduct}[1]%
   {\ensuremath{\mathcal{T} \{ \, #1 \, \} } }
\newcommand{\aTproduct}[1]%
   {\ensuremath{\mathcal{T}^{\ast} \{ \, #1 \, \} } }
\newcommand{\Nproduct}[1]%
   {\ensuremath{\mathcal{N} \{ \, #1 \, \} } }
\newcommand{\ctpTproduct}[1]%
   {\ensuremath{\mathcal{T}_{\mathcal{C}} \{ \, #1 \, \} }}
\newcommand{\tauordered}[1]%
   {\ensuremath{\mathcal{T}_{\tau} \{ \, #1 \, \} }}
\newcommand{\expectTproduct}[1]%
   {\ensuremath{\langle \, \mathcal{T} \{ \, #1 \, \} \, \rangle}}
\newcommand{\expectTCproduct}[1]%
   {\ensuremath{\langle \, \mathcal{T}_{\mathcal{C} \{ \, #1 \, \} \, \rangle}}}
\newcommand{\expectComm}[2]%
   {\ensuremath{\langle \, [ \, #1 , #2 \, ] \, \rangle}}
\newcommand{\expectPbracket}[2]%
   {\ensuremath{\langle \, \{ \, #1 , #2 \, \} \, \rangle}}
\newcommand{\threej}[6]%
{\begin{pmatrix} #1 & #2 & #3 \\ #4 & #5 & #6 \end{pmatrix}}
\newcommand{\sixj}[6]%
{\begin{Bmatrix} #1 & #2 & #3 \\ #4 & #5 & #6 \end{Bmatrix}}
\newcommand{\ninej}[9]%
{\begin{Bmatrix} #1 & #2 & #3 \\ #4 & #5 & #6 \\%
 #7 & #8 & #9 \end{Bmatrix}}
\newcommand{\reducedme}[3]%
{\langle \, #1 \, \Vert \, #2 \, \Vert \, #3 \, \rangle }
\begin{document}
%
%%%%%%%%%%%%%%%%%%%%%%%%%%%%%%%%%%%%%%%%%%%%%%%%%%%%%%%%%%%%%%%%%%%%%%
%
% titlepage
%
%\begin{figure}[!t]
%\vskip -1.3cm
%\leftline{ \includegraphics[width=1.1cm]{Images/UNH-logo.pdf} \qquad 
%           \includegraphics[width=3cm]{Images/SFI-logo-2.jpeg} 
%           \includegraphics[width=3cm]{Images/LANL-logo-4.jpeg} }
%\end{figure}
%
%
\preprint{LA-UR-11-01956}
\title[BCS-BEC crossover]
   {Auxiliary field formalism for dilute fermionic atom gases with tunable interactions}

\author{Bogdan Mihaila}
\email{bmihaila@lanl.gov}
\affiliation{% Materials Science and Technology Division,
   Los Alamos National Laboratory,
   Los Alamos, NM 87545}

\author{John F. Dawson}
\email{john.dawson@unh.edu}
\affiliation{Department of Physics,
   University of New Hampshire,
   Durham, NH 03824}

\author{Fred Cooper}
\email{cooper@santafe.edu}
\affiliation{% Theoretical Division,
   Los Alamos National Laboratory,
   Los Alamos, NM 87545}
\affiliation{Santa Fe Institute,
   Santa Fe, NM 87501}

\author{Chih-Chun Chien}
\email{chinchun@lanl.gov}
\affiliation{% Theoretical Division,
   Los Alamos National Laboratory,
   Los Alamos, NM 87545}

\author{Eddy Timmermans}
\email{eddy@lanl.gov}
\affiliation{% Center for Nonlinear Studies,
   Los Alamos National Laboratory,
   Los Alamos, NM 87545}

%\date{\today, \now \ EST}


\begin{abstract}
We develop the auxiliary field formalism corresponding to a dilute system of spin-1/2 fermions. 
This theory represents the Fermi counterpart of the BEC theory developed recently by F.~Cooper \emph{et al.} [Phys. Rev. Lett.  \textbf{105}, 240402 (2010)] to describe a dilute gas of Bose particles.
Assuming tunable interactions, this formalism is appropriate for the study of  the  crossover from the regime of Bardeen-Cooper-Schriffer (BCS) pairing to the regime of Bose-Einstein condensation (BEC) in ultracold fermionic atom gases.  We show that when applied to the Fermi case at zero temperature, the leading-order auxiliary field (LOAF) approximation gives the same equations as those obtained in the standard BCS variational picture. At finite temperature, LOAF leads to the theory discussed by by S\'a de~Melo, Randeria, and Engelbrecht [Phys. Rev. Lett. \textbf{71}, 3202(1993); Phys. Rev. B \textbf{55}, 15153(1997)]. As such, LOAF provides a unified framework to study the interacting Fermi gas. 
%Furthermore, we show that LOAF satisfies Tan's relation regarding the momentum distribution of fermions at asymptotically large momenta and the ``adiabatic sweep'' theorem. 
The mean-field results discussed here can be systematically improved upon by calculating the one-particle irreducible (1-PI) action corrections, order by order.
\end{abstract}
\maketitle
%\tableofcontents
%\listoffigures

%
%%%%%%%%%%%%%%%%%%%%%%%%%%%%%%%%%%%%%%%%%%%%%%%%%%%%%%%%%%%%%%%%%%%%%%
%
\section{\label{s:intro}Introduction}

One of the most remarkable achievements of the past decade concerns the advent of experimental studies in ultracold fermionic atom gases of the crossover from the regime of Bardeen-Cooper-Schriffer (BCS) weakly bound Cooper pairs to the regime of Bose-Einstein condensation (BEC) of diatomic molecules~\cite{r:OHara:2002ly, r:Gehm:2003ve, r:Gehm:2003qf, r:Bourdel:2003bh, r:Regal:2003dq, r:Regal:2003cr, r:Regal:2004nx, r:Gupta:2003oq, r:Gupta:2004kl, r:Zwierlein:2003tg, r:Bartenstein:2004hc}. In turn, these studies made possible for the first time the experimental study of the ground-state properties of a many-body system composed of spin-1/2 fermions interacting via a zero-range, infinite scattering length contact interaction. This regime is known as the unitarity limit~\cite{r:unitarity} and is of particular interest in astrophysics because of its implications regarding the equation of state for neutron matter \cite{r:Heiselberg:2000zr}, thus emphasizes the far-reaching implications of these recent studies.

Much of the theoretical work on systems composed of spin-1/2 fermions interacting via an adjustable, attractive potential has focussed on interactions that are governed by a single parameter, namely the s-wave scattering length, $a$, of two atoms with different spin components~\cite{r:Leggett:1980fk,r:Randeria:1995ij}.  This description is valid only if $|a| \gg r_0$ and $k_F \, r_0 \ll 1$, where $r_0$ is the range of the potential and $k_\mathrm{F}$ denotes the Fermi momentum of the gas and is conventionally related to the total density of particles, $\rho$, by the noninteracting Fermi gas formula
\begin{equation*}
   \rho 
   =
   \sum_{\sigma}
   \int_0^{\kF} \!\!\! \frac{\rd^3 k}{(2\pi)^3}
   =
   \frac{\kF^3}{3 \pi^2} \>,
\end{equation*}
The above momentum integral is performed over the interior volume of the Fermi sphere and $\sigma$ denotes the spin component of the fermion, i.e. $\sigma = \pm 1/2$.  Then, the only independent dimensionless variable in the problem is $a \, k_F$. Thus, this description is only applicable to dilute systems like the ultracold fermionic atom gases, not the high-density regime found in conventional superconductors \cite{r:Schrieffer:1964bs}. The potential supports a 2-body bound state for $(a \, k_F)^{-1}>0$, but this molecular state passes through zero energy and vanishes into the continuum at $(a \, k_F)^{-1} = 0$, the position of the Feshbach resonance. Then, the BCS and BEC limits correspond to $(a \, k_F)^{-1}$$\rightarrow$$- \infty$ and $(a \, k_F)^{-1}$$\rightarrow$$+\infty$, respectively, whereas the unitarity limit is defined as the limit near Feshbach resonances where $a$ is much larger than the inter-particle distance ($|a| \, \kF \gg 1$) and corresponds to the BCS to BEC crossover at the singularity of the scattering length.  This limit is the same when approached with positive or negative scattering length. In the unitarity limit, the correlations are deemed to be significant and the system attains universal behavior, independent of the shape of the potential and dependent only on the particle density and the system dimensionality~\cite{r:Nishida:2006fk,r:Nishida:2007uq,r:Mihaila:2009kx}.

The importance of correlations in the ground state of dilute fermionic matter in the unitarity limit is measured by the numerical value of the ratio, $\varepsilon / \varepsilon_0$, where $\varepsilon$ and $\varepsilon_0$ denote the ground-state energies per particle of the interacting and noninteracting systems, respectively. We recall the noninteracing energy density is defined as \begin{equation*}
   \varepsilon_0 
   =
   \frac{1}{\rho} \,
   \sum_{\sigma}
   \int_0^{\kF} \!\!\! \frac{\rd^3 k}{(2\pi)^3} \ \epsilon_k
   =
   \frac{3}{5} \,  \epsilonF \>,
\end{equation*}
where we introduced the notation,
$\epsilon_k = \gamma k^2$ with $\gamma = \hbar^2 / (2m)$, and $\epsilonF = \gamma \kF^2$ is  the Fermi energy.

An upper bound to the value of $\varepsilon / \varepsilon_0$  at zero temperature was set by the quantum Monte Carlo (QMC) study performed by 
Carlson \emph{et al}.~\cite{r:Carlson:2008fk,r:Carlson:2005uq,r:Chang:2004kx,r:Carlson:2003fv} that gave the value $(\varepsilon / \varepsilon_0)_\mathrm{QMC} = 0.40(01)$.  Instead, the ``universal'' curve describing the BCS to BEC crossover in the standard BCS variational picture~\cite{r:Engelbrecht:1997fk,r:Parish:2005fk} derived by Leggett~\cite{r:Leggett:1980fk} gives the BCS value of the chemical potential to the Fermi energy ratio, $(\mu / \epsilon_F)_{BCS} = 0.59$. 
Other theoretical and experimental values for $\varepsilon/ \varepsilon_0$ are summarized elsewhere~\cite{r:Heiselberg:2004dz,r:Levinsen:2006kl,r:Burovski:2006qa,r:Nikolic:2007fu}. 

Recently we introduced  a new theoretical framework for a dilute gas of Bose particles with tunable interactions~\cite{r:Cooper:2010fk}, the bosonic counterpart of the system of fermions discussed in this paper. For the Bose system, our theoretical description is based on a loop expansion of the one-particle irreducible (1-PI) effective action in terms of composite-field propagators by rewriting the Lagrangian in terms of auxiliary fields related to the normal and anomalous densities~\cite{r:Cooper:2010fk}. The leading-order auxiliary field (LOAF) approximation in the case of an interacting dilute Bose gas describes a large interval of values of the coupling constant, satisfies Goldstone's theorem and yields a Bose-Einstein transition that is second order, while also predicting reasonable values for the depletion.  

In this paper we will  derive the corresponding auxiliary field formalism for a dilute fermionic atom gas with tunable interactions, thus establishing the generality of our auxiliary field formalism. We will show that the LOAF approximation in the fermionic case corresponds to the BCS ansatz.  At zero temperature the fermionic LOAF equations are the same as the equations derived by Leggett~\cite{r:Leggett:1980fk}, whereas the finite-temperature results correspond to those discussed earlier by S\'a de~Melo, Randeria, and Engelbrecht~\cite{r:Melo:1993vn,r:Engelbrecht:1997fk}. Hence, we find that the BCS ansatz is the only relevant auxiliary field theory in a dilute interacting Fermi gas.  In the auxiliary field approach, one can systematically improve upon the LOAF approximation by  calculating the 1-PI action corrections, order by order.
A related approach for the relativistic four-fermi theory can be found in Refs.~\onlinecite{r:Chodos:2000fk,r:CCMS01,r:Mihaila:2006fk}. 

This paper is organized as follows: In Sec.~\ref{s:thm} we discuss the partition function for an infinite homogeneous system of spin-1/2 fermions with arbitrary populations of spin-up and spin-down fermions. In Sec.~\ref{s:auxformI} we discuss rewriting the Lagrangian in terms of auxiliary fields. The resulting effective action is discussed in Sec.~\ref{s:Seff}. The corresponding properties of an uniform system in equilibrium are derived in Sec.~\ref{s:thermalequib}. In Sec.~\ref{ss:caseA}, we specialize to the case of systems with equal populations of spin-up and spin-down fermions. We summarize our findings in Sec.~\ref{s:concl}.

%
%%%%%%%%%%%%%%%%%%%%%%%%%%%%%%%%%%%%%%%%%%%%%%%%%%%%%%%%%%%%%%%%%%%%%%
%
\section{\label{s:thm}The partition function and path integrals}

For a grand canonical ensemble, the partition function for a collection of interacting Fermi particles can be written as
\begin{equation}\label{thm.e:rhoGCEI}
   Z[T,\mu,V]
   =
   e^{- \beta \, \Omega[T,\mu,V] }
   =
   \Tr{ e^{- \beta \, ( \, \hH - \mu \, \hN \, ) } } \>,
\end{equation}
where $\Omega[T,\mu,V]$ is the grand potential and we have set $\beta = 1 / T$, with the temperature measured in $\kBolt$ units.  Here $T$ is the temperature, $\mu$ the chemical potential, and $V$ the volume.  Using the second law of thermodynamics, we find 
\begin{align}
   N[T,\mu,V]
   &=
   - \Partial{\,\Omega}{\mu}{T}{V} \>,
   \notag \\
   p[T,\mu,V]
   &=
   - \Partial{\,\Omega}{V}{T}{\mu} \>,
   \label{thm.e:SNpI} \\
   S[T,\mu,V]
   &=
   - \Partial{\,\Omega}{T}{\mu}{V} \>,
   \notag
\end{align}
%Now since the energy $E$ is given by
and the energy $E$ is given by
\begin{equation}\label{thm.e:U}
   E
   =
   \Omega + T S + \mu N \>.
\end{equation}
%the isentropic heat capacity at constant $\mu$ and $V$ is given by
%\begin{align}
%   C_{\mu,V}
%   &=
%   \Partial{\,E}{T}{\mu}{V}
%   =
%   T \,
%   \Partial{\,S}{T}{\mu}{V}
%   +
%   \mu \,
%   \Partial{\,N}{T}{\mu}{V}
%   \label{thm.e:cS} \\
%   &=
%   - 
%   T \,
%   \Bigl ( \frac{\partial^2 \, \Omega}{\partial T^2} \Bigr )_{\mu,V}
%   -
%   \mu \,
%   \Bigl ( \frac{\partial^2 \, \Omega}{\partial T \, \partial \mu} \Bigr )_{V}
%   > 0 \>.
%   \notag
%\end{align}
%The compressibility $\kappa_{T,V}$ at constant $T$ and $V$ is given by
%\begin{equation}\label{thm.e:numfluxII}
%   \kappa_{T,V}
%   =
%   \frac{1}{\beta} \,
%   \Partial{N}{\mu}{T}{V} 
%   =
%   -
%   \frac{1}{\beta} \,
%   \Bigl ( 
%      \frac{\partial^2 \, \Omega}{\partial \mu^2}
%   \Bigr )_{\! T,V}
%   > 0 \>.
%\end{equation}
The partition function can be written as a path integral,
\begin{equation}\label{pf.e:Z-I}
   Z[T,\mu,V]
   =
   \calN
   \iint \rD \psi \, \rD \psi^{\ast} \, 
      e^{-S[\psi,\psi^{\ast};T,\mu,V]} \>,
\end{equation}
where $S[\psi,\psi^{\ast};T,\mu,V]$ the negative of the Euclidian (thermal) action obtained by mapping the physical action to imaginary time, $t \mapsto - i \hbar \tau$.  We consider here the action $\calS$ for a collection of fermions interacting by means of a short-range contact potential $V(r) = \lambda_0 \, \delta(\br)$ is given by
\begin{equation}\label{pf.e:Sdef}
   \calS[ \, \psi,\psi^{\ast} \, ]
   =
   \int [\rd x] \,
   \calL[ \, \psi,\psi^{\ast} \, ] \>,
\end{equation}
where we have put $\int \! \rd^3 x \int_{0}^{\beta} \! \rd \tau$, and where
\begin{align}
   \calL[ \, \psi,\psi^{\ast} \, ] 
   &=
   \sum_{\sigma}
   \biggr \{ \,
      \frac{1}{2} \,
      \Bigl [ \,
         \psi_{\sigma}^{\ast}(x) \, 
         \frac{\partial \psi_{\sigma}(x)}{\partial \tau}
         +
         \psi_{\sigma}(x) \,
         \frac{\partial \psi_{\sigma}^{\ast}(x)}{\partial \tau} \, 
      \Bigr ]
      \notag \\
      & \qquad
      +
      \psi^{\ast}_{\sigma}(x) \, 
      \Bigl [ \,
         -
         \gamma \nabla^2
         -
         \mu_{\sigma} \,   
      \Bigr ] \, 
      \psi^{\phantom\ast}_{\sigma}(x)
      \label{pf.e:LagI} \\
      & \qquad
      +
      \frac{\lambda_0}{2} \,
      \psi^{\ast}_{\sigma}(x) \, \psi^{\ast}_{-\sigma}(x) \,
      \psi^{\phantom\ast}_{-\sigma}(x) \, \psi^{\phantom\ast}_{\sigma}(x) \,
   \biggr \} \>.
   \notag
\end{align}
We have suppressed the dependence of quantities on the thermodynamic variables $(T,\mu_{\pm},V)$.  The fields are described by two-component complex anticommuting Grassmann fields $\psi_{\sigma}(x)$ which obey the algebra,
\begin{gather}
   \AntiComm{\psi^{\phantom\ast}_{\sigma}(x)}
            {\psi^{\ast}_{\sigma'}(x')}
   =
   \AntiComm{\psi^{\phantom\ast}_{\sigma}(x)}
            {\psi^{\phantom\ast}_{\sigma'}(x')}
   \label{pf.e:psialgebra} \\
   =
   \AntiComm{\psi^{\ast}_{\sigma}(x)}
            {\psi^{\ast}_{\sigma'}(x')}
   =
   0 \>,
   \notag
\end{gather}
with $x \equiv \Set{\br,\tau}$, and $\sigma = \pm 1$ correspond to the usual spin-up ($\uparrow$) and spin-down ($\downarrow$) fermions.  Using a \emph{left} derivative convention for Grassmann derivatives, variation of the action with respect to $\psi^{\ast}_{\sigma}(x)$ leads to thermal equations of motion,
\begin{equation*}\label{pf.e:psieomI}
   \Bigl \{ \,
      -
      \gamma \nabla^2
      +
      \frac{ \partial }{\partial \tau}
      -
      \mu_{\sigma}
      +
      \lambda_0 \, 
      \bigl [ \,
          \psi^{\ast}_{-\sigma}(x) \, \psi^{\phantom\ast}_{-\sigma}(x) \,
       \bigr ] \,
    \Bigr \} \,
    \psi_{\sigma}(x)
    =
    0 \>.
\end{equation*}
(No sum over $\sigma$ here.)  
Particle densities $\rho_{\sigma}^{\phantom\ast}(x)$ and $\kappa_{\sigma}^{\phantom\ast}(x)$ are defined by
\begin{subequations}\label{pf.e:nadensdef}
\begin{align}
   \rho_{\sigma}^{\phantom\ast}(x)
   &=
   \psi^{\ast}_{\sigma}(x) \, \psi^{\phantom\ast}_{\sigma}(x)
   =
   \rho_{\sigma}^{\ast}(x) \>,
   \label{fp.e:rhodendef} \\
   \kappa_{\sigma}(x)
   &=
   \psi^{\phantom\ast}_{\sigma}(x) \, 
   \psi^{\phantom\ast}_{-\sigma}(x)
   =
   - \kappa_{-\sigma}(x) \>.
   \label{fp.e:kappadendef}
\end{align}
\end{subequations}
which have the property that
\begin{subequations}\label{pf.e:denprops}
\begin{align}
   \sum_{\sigma}
   \rho^{\phantom\ast}_{\sigma}(x) \, \rho^{\ast}_{-\sigma}(x)
   &=
   \sum_{\sigma}
   \rho^{\ast}_{\sigma}(x) \, \rho^{\phantom\ast}_{-\sigma}(x)
   \label{pf.e:rhoprop} \\
   &=
   +
   \sum_{\sigma}
   \psi^{\ast}_{\sigma}(x) \, 
   \psi^{\ast}_{-\sigma}(x) \, 
   \psi^{\phantom\ast}_{-\sigma}(x) \,
   \psi^{\phantom\ast}_{\sigma}(x) \>,
   \notag \\
   \sum_{\sigma}
   \kappa_{\sigma}^{\phantom\ast}(x) \,
   \kappa_{-\sigma}^{\ast}(x)
   &=
   \sum_{\sigma}
   \kappa_{\sigma}^{\ast}(x) \,
   \kappa_{-\sigma}^{\phantom\ast}(x)
   \label{pf.e:kappadenprop} \\
   &=
   -
   \sum_{\sigma}
   \psi^{\ast}_{\sigma}(x) \, 
   \psi^{\ast}_{-\sigma}(x) \, 
   \psi^{\phantom\ast}_{-\sigma}(x) \,
   \psi^{\phantom\ast}_{\sigma}(x) \>.
   \notag
\end{align}
\end{subequations}
We define $\kappa(x) \equiv \kappa_{+}(x) = - \kappa_{-}(x)$.  
The densities $\rho_{\pm}(x)$ are real and independent whereas $\kappa(x)$ is complex.  Introducing four component basis vectors $\psi^{a}(x)$ and $\psi_{a}(x) $,
\begin{align}
   \psi^{a}(x)
   &=
   \Set{ 
      \psi^{\phantom\ast}_{+}(x), 
      \psi^{\phantom\ast}_{-}(x), 
      \psi^{\ast}_{+}(x), 
      \psi^{\ast}_{-}(x)
       } \>,
   \label{pf.e:psiu} \\
   \psi_{a}(x)
   &=
   \Set{ 
      \psi^{\ast}_{+}(x), 
      \psi^{\ast}_{-}(x), 
      \psi^{\phantom\ast}_{+}(x), 
      \psi^{\phantom\ast}_{-}(x)
       } \>,
   \notag   
\end{align}
the density matrix can then be written as
\begin{align}
   \rho_a{}^b(x)
   &=
   \psi_{a}(x) \, \psi^{b}(x)
   \label{pf.e:denmat} \\
   &=
   \begin{pmatrix}
      \rho_{+}(x) & 0 & 0 & - \kappa^{\ast}(x) \\
      0 & \rho_{-}(x) & \kappa^{\ast}(x) & 0 \\
      0 & \kappa(x) & - \rho_{+}(x) & 0 \\
      -\kappa(x) & 0 & 0 & - \rho_{-}(x)
   \end{pmatrix} \>.
   \notag
\end{align}

%
%%%%%%%%%%%%%%%%%%%%%%%%%%%%%%%%%%%%%%%%%%%%%%%%%%%%%%%%%%%%%%%%%%%%%%
%
\section{\label{s:auxformI}Auxiliary fields}

Following the Bose case \cite{r:Cooper:2010fk}, we use the Hubbard-Stratonovitch transformation \cite{r:Hubbard:1959kx,*r:Stratonovich:1958vn} to introduce auxiliary fields for each of the densities described above in order to eliminate the quadratic interaction term in the Lagrangian in favor of cubic interactions between the Fermi field and the auxiliary fields.  There are six independent auxiliary fields possible in our case, two real fields $\chi_{\sigma}(x)$, and two complex fields $\Delta_{\sigma}(x)$, corresponding to the densities $\rho_{\sigma}(x)$ and $\kappa_{\sigma}(x)$ respectively.
The auxiliary Lagrangian is defined by
\begin{equation}\label{AF.e:Lauxdef}
   \calL_{\text{aux}}[\psi,\chi,\Delta]
   =
   -
   \calL_{\chi}[\psi,\chi]
   +
   \calL_{\Delta}[\psi,\Delta] \>,
\end{equation}
where
\begin{subequations}\label{AF.e:Ldefs}
\begin{align}
   \calL_{\chi}[\psi,\chi]
   &=
   \frac{1}{2\lambda_0} \,
   \sum_{\sigma}
   \bigl [ \,
      \chi_{\sigma}^{\phantom\ast}(x)
      -
      \lambda_0 \, \rho_{\sigma}(x) \, \sin\theta \,
   \bigr ] 
   \label{AF.e:Lauxchi} \\
   & \qquad\qquad
   \times
   \bigl [ \,
      \chi_{-\sigma}(x)
      -
      \lambda_0 \, \rho_{-\sigma}(x) \, \sin\theta \,
   \bigr ]
   \notag \\
   &=
   \frac{1}{2\lambda_0} \,
   \sum_{\sigma} 
   \chi_{\sigma}(x) \, \chi_{-\sigma}(x)
   \notag \\
   & \quad
   -
   \frac{\sin\theta}{2}
   \sum_{\sigma} 
   \bigl [ \,
      \chi_{\sigma}(x) \, 
      \rho_{-\sigma}(x)
      +
      \rho_{\sigma}(x) \, 
      \chi_{-\sigma}(x) \, 
   \bigr ] 
   \notag \\
   & \quad
   +
   \frac{\lambda_0 \, \sin^2\theta}{2} \,
   \sum_{\sigma}
   \psi^{\ast}_{\sigma}(x) \, 
   \psi^{\ast}_{-\sigma}(x) \, 
   \psi^{\phantom\ast}_{-\sigma}(x) \,
   \psi^{\phantom\ast}_{\sigma}(x) \>,
\end{align}
and
\begin{align}
   \calL_{\Delta}[\psi,\Delta]
   &=
   \frac{1}{2\lambda_0} \,
   \sum_{\sigma}
   \bigl [ \,
      \Delta_{\sigma}^{\phantom\ast}(x)
      -
      \lambda_0 \, \kappa_{\sigma}(x) \, \cos\theta \,
   \bigr ] \\
   & \qquad\qquad
   \times
   \bigl [ \,
      \Delta_{-\sigma}^{\ast}(x)
      -
      \lambda_0 \, \kappa_{-\sigma}^{\ast}(x) \, \cos\theta \,
   \bigr ]
   \notag \\
   &=
   \frac{1}{2\lambda_0} \,
   \sum_{\sigma} 
   \Delta_{\sigma}^{\phantom\ast}(x) \, \Delta_{-\sigma}^{\ast}(x)
   \notag \\
   & \quad
   -
   \frac{\cos\theta}{2}
   \sum_{\sigma} 
   \bigl [ \,
      \Delta_{\sigma}^{\phantom\ast}(x) \, 
      \kappa_{-\sigma}^{\ast}(x)
      +
      \kappa_{\sigma}^{\phantom\ast}(x) \, 
      \Delta_{-\sigma}^{\ast}(x) \, 
   \bigr ] 
   \notag \\
   & \quad
   -
   \frac{\lambda_0 \, \cos^2\theta}{2} \,
   \sum_{\sigma}
   \psi^{\ast}_{\sigma}(x) \, 
   \psi^{\ast}_{-\sigma}(x) \, 
   \psi^{\phantom\ast}_{-\sigma}(x) \,
   \psi^{\phantom\ast}_{\sigma}(x) \>.
   \notag   
\end{align}
\end{subequations}
Here we have introduced an angle $\theta$.  The auxiliary fields obey the same properties as the corresponding densities, so we define $\Delta(x) \equiv \Delta_{+}(x) = - \Delta_{-}(x)$.  
So \emph{adding} $\calL_{\text{aux}}$ to $\calL$ given in Eq.~\eqref{pf.e:LagI}, eliminates the four-point Fermi interaction.  Using the basis vectors given in Eq.~\eqref{pf.e:psiu}, the action can be written in a compact way as
\begin{align}
   &\calS[\Psi,J] 
   =
   \frac{1}{2} \,
   \iint [\rd x] \, [\rd x'] \,
   \psi_{a}(x) \, 
   \calG^{-1}{}^{a}{}_{b}[\phi](x,x') \, 
   \psi^{b}(x')
   \notag \\
   & \quad
   +
   \int [\rd x] \,
   \Bigl [ \,
      -
      \frac{ (\phi_i(x) + \mu_i) \,
             g^i{}_j(\theta) \, 
             (\phi^j(x) + \mu^j) }
           { 2 \, \lambda_0 }
      \notag \\
      & \qquad\qquad
      +
      \psi_{a}(x) \, g^a{}_b \, j^b(x)
      +
      \phi_i(x) S^{i}(x) \,
   \Bigr ] \>,
   \label{AF.e:action}
\end{align}
where the inverse Green function is given by
\begin{equation}\label{AF.e:calGdef}
   \calG^{-1}{}^{a}{}_{b}[\phi](x,x') 
   = 
   \delta(x,x') \,
   \bigl [ \,
      \calG^{-1}_{0}{}^{a}{}_{b}[\phi]
      +
      \calV^{a}{}_b[\phi](x) \,
   \bigr ]
\end{equation}
with
\begin{equation}\label{AF.e:calGzerodef}
   \calG^{-1}_{0}{}^{a}{}_{b}[\phi] 
   = 
   \begin{pmatrix}
      h_{+} & 0 & 0  & 0 \\
      0 & h_{+} & 0  & 0 \\
      0 & 0 & -h_{-} & 0 \\
      0 & 0 & 0 & -h_{-}
   \end{pmatrix} \>,
\end{equation}
where we have defined $h_{\pm}$ as the operators
\begin{equation}\label{AF.e:hdefs}
   h_{\pm}
   =
   -
   \gamma \nabla^2
   \pm
   \frac{\partial}{ \partial \tau } \>,
\end{equation}
and
\begin{equation}\label{AF.e:calVdef}
   \calV^{a}{}_b[\phi](x)
   =
   \begin{pmatrix}
      \chi'_{-}(x) & 0 & 0 & -\Delta'(x) \\
      0 & \chi'_{+}(x) & \Delta'(x) & 0 \\
      0 & \Delta^{\prime\,\ast}(x) & - \chi'_{-}(x) & 0 \\
      -\Delta^{\prime\,\ast}(x) & 0 & 0 & - \chi'_{+}(x) 
   \end{pmatrix} \>,
\end{equation}
Here we have redefined the anomalous fields by setting
\begin{gather}
   \chi'_{\pm}(x) 
   \equiv
   \chi_{\pm}(x) \, \sin\theta - \mu_{\mp} \>,
   \label{AF.e:chiptchipDeltapdefs} \\
   \Delta'(x) 
   \equiv
   \Delta_{+}(x) \, \cos\theta
   = 
   - \Delta_{-}(x) \, \cos\theta \>,
   \notag
\end{gather}
and introduced four component anomalous fields $\phi^i(x)$, a constant vector $\mu^i$, and currents $S^i(x)$ by the definitions
\begin{align}
   \phi^{i}(x)
   &=
   \Set{ \chi'_{+}(x), \chi'_{-}(x), \Delta'(x), \Delta^{\prime\,\ast}(x) } \>,
   \label{ea.e:AFASupperdef} \\
   \mu^i_0
   &=
   \Set{ \, \mu_{-}, \mu_{+},0,0 } \>,
   \notag \\
   S^{i}(x)
   &=
   \Set{ s_{+}(x), s_{-}(x), S(x), S^{\ast}(x) } \>,
   \notag
\end{align}
with
\begin{align}
   \phi_{i}(x)
   &=
   \Set{ \chi'_{-}(x), \chi'_{+}(x), \Delta^{\prime\,\ast}(x), \Delta'(x) } \>,
   \label{ea.e:AFASlowerdef} \\
   \mu_{0\,i}
   &=
   \Set{ \, \mu_{+}, \mu_{-},0,0 } \>,
   \notag \\
   S_{i}(x)
   &=
   \Set{ s_{-}(x), s_{+}(x), S^{\ast}(x), S(x) } \>.
   \notag
\end{align}
The tensors $g^a{}_b$ and $g^i{}_j(\theta)$ are defined by
\begin{align}
   g^a{}_b
   &=
   \Diag{ 1, 1, -1, -1 } \>,
   \label{ea.e:ZabIijdef} \\
   g^i{}_j(\theta)
   &=
   \Diag{ 1/\sin^2\theta, 1/\sin^2\theta, 1/\cos^2\theta, 1/\cos^2\theta } \>.
   \notag
\end{align}
The Grassmann fields $\psi^{a}(x)$ and $\psi_{a}(x)$ are defined as in Eq.~\eqref{pf.e:psiu}.  It will be useful for notational purposes to also define ten component fields and currents using Greek indices as
\begin{align}
   \Psi^{\alpha}(x)
   &=
   \Set{ \psi^{a}(x),\phi^{i}(x) } \>,
   \label{ea.e:PhiJdefs} \\
   J^{\alpha}(x)
   &=
   \Set{ j^{a}(x), S^i(x) } \>.
   \notag
\end{align}
Note that the $\psi^a(x)$ fields are Grassmann fields whereas the $\phi^{i}(x)$ fields are commuting fields, so the vectors $\Psi^{\alpha}(x)$ and $J^{\alpha}(x)$ are superfields.

Setting $\chi' = 0$ and then $\theta = 0$ so that only $\Delta$ survives, recovers the first order S\'a de~Melo-Randeria-Engelbrecht theory\cite{r:Melo:1993vn}.

%
%%%%%%%%%%%%%%%%%%%%%%%%%%%%%%%%%%%%%%%%%%%%%%%%%%%%%%%%%%%%%%%%%%%%%%
%
\section{\label{s:Seff}Effective action}

The partition function is now given by a path integral over all fields,
\begin{align}
   Z[\, J \,]
   &=
   e^{-\beta \, \Omega[\, J \,]}
   \label{ea.e:Z-I} \\
   &=
   \calN \iint
   \prod_a \rD \psi^a \, \rD \psi_a \, 
   \!\! \iint
   \prod_i \rD \phi^i \, \rD \phi_i \,
   e^{ - \calS[\, \Psi,J \,]} \>,
   \notag
\end{align}
where $\calS[\, \Psi,J \,]$ is given in Eq.~\eqref{AF.e:action} and again we have suppressed the dependence of $Z$, $\Omega$, and $\calS$ on the thermodynamic variables.  The thermodynamic partition function is found by setting the currents to zero.  Thermal average values of the fields are given by
\begin{align}
   \Expect{\psi^a(x)}
   &=
   \frac{-1}{Z} \, 
   \frac{\delta Z}{\delta j_a(x)} \, \Big |_{j=S=0}
   =
   \beta \frac{\delta \Omega}{\delta j_a(x)}  \, \Big |_{j=S=0} \>,
   \label{ea.e:avefields} \\
   \Expect{\phi^i(x)}
   &=
   \frac{-1}{Z} \, 
   \frac{\delta Z}{\delta S_i(x)} \, \Big |_{j=S=0}
   =
   \beta \frac{\delta \Omega}{\delta S_i(x)} \, \Big |_{j=S=0} \>.
   \notag   
\end{align}
Since the action is now quadratic in the fields $\psi^a(x)$, we can integrate these out and obtain an effective action,
\begin{equation}\label{ea.e:Z-II}
   Z[\, J \,]
   =
   \calN \iint
   \prod_i \rD \phi^i \, \rD \phi_i \,
   e^{ - \calS_{\text{eff}}[\, \phi,J \,]} \>,
\end{equation}
where the effective action $\calS_{\text{eff}}[\, \phi,J \,]$ is given by
\begin{align}
   &\calS_{\text{eff}}[\, \phi,J \,]
   =
   \frac{1}{2}
   \iint  [\rd x ]\, [ \rd x' ] \,
   j_a(x) \, g^a{}_b \, \calG^b{}_c[\phi](x,x') \, j^c(x')
   \notag \\
   & \qquad
   - 
   \frac{1}{2} \,
   \int [\rd x] \,
   \mathrm{Tr}
   \bigl [ \,
      \Ln{  \calG^{-1}[\phi](x,x) } \,
   \bigr ] 
   \notag \\
   &
   +
   \int [\rd x] \,
   \Bigl [ \, 
      -
      \frac{ ( \phi_i(x) + \mu_{0\,i} ) \,
             g^i{}_j \, 
             ( \phi^j(x) + \mu_0^j ) }
           { 2 \, \lambda_0 }
      +
      \phi_i(x) S^{i}(x) \,
   \Bigr ] \>.
   \label{ea.e:Seff}
\end{align}
We evaluate the remaining path integral by expanding the effective action about a point $\phi_0^i$,
\begin{align}
   &\calS_{\text{eff}}[\, \phi,J \,]
   =
   \calS_{\text{eff}}[\, \phi_0,J \,]
   \label{ea.e:Sexpand} \\
   & \qquad
   +
   \int [\rd x] \,
   \frac{\delta \calS_{\text{eff}}[\, \phi,J \,]}
        {\delta \phi^i(x)} \Big |_{\phi_0} \!\!
   ( \, \phi^i(x) - \phi_0^{i}(x) \, )
   \notag \\
   & \quad
   +
   \frac{1}{2} 
   \iint [\rd x] \, [\rd x'] \,
   \frac{\delta^2 \calS_{\text{eff}}[\, \phi,J \,]}
        {\delta \phi^i(x) \, \delta \phi^j(x')} \Big |_{\phi_0} \!\!
   \notag \\
   & \qquad\qquad
   \times
   ( \, \phi^i(x) - \phi_{0}^{i}(x) \, ) \,
   ( \, \phi^j(x') - \phi_{0}^{j}(x') \, )
   +
   \dotsb
   \notag
\end{align}
and evaluating the path integral by the method of steepest descent.  The vanishing of the first derivatives define the saddle point $\phi_0^i$, which gives
\begin{align}
   &\frac{ g^i{}_j \, [ \, \phi_0^j(x) + \mu^j \, ] }{ \lambda_0 }
   =
   \rho_0^{i}[\phi_0,J](x)
   \label{ea.e:statpt} \\
   & \qquad\qquad\qquad
   -
   \frac{1}{2} \, 
   \mathrm{Tr}
   \bigl [ \,
      \calV^i \, \calG[\phi_0](x,x) \,
   \bigr ] 
   +
   S^{i}(x) \>.
   \notag
\end{align}
Here we have used
\begin{equation*}\label{ea.e:GinvG}
   \int [\rd x'] \,
   \calG^{-1}{}^a{}_b[\phi](x,x') \,
   \calG^b{}_c[\phi](x',x'')
   =
   g^a{}_c \, \delta(x - x'') \>,
\end{equation*}
so that
\begin{align}
   &\frac{\delta \calG^a{}_e[\phi](x_1,x_4) }
        { \delta \phi_i(x) }
   =
   -
   \iint [\rd x_2] \, [\rd x_3] \,
   g^a{}_b \,
   \calG^{b}{}_{c}[\phi](x_1,x_2) \,
   \notag \\
   & \qquad\qquad
   \times
   \frac{\delta \calG^{-1\,c}{}_{d}[\phi](x_2,x_3) }
        { \delta \phi_i(x) } \,
   \calG^{d}{}_{e}[\phi](x_3,x_4)
   \notag \\
   &=
   -
   g^{a}{}_{b} \,
   \calG^b{}_c[\phi](x_1,x) \,
   \calV^{i\,c}{}_d \, 
   \calG^d{}_e[\phi](x,x_4) \>,
   \label{ea.e:dGdchi}
\end{align}
and defined the constant matrices $\calV^{i\,a}{}_b$ by
\begin{equation}\label{ea.e:calVidefs}
   \frac{\delta \calG^{-1\,a}{}_b[\phi](x_2,x_3) }
        { \delta \phi_i(x) }
   =
   \calV^{i\,a}{}_b \, \delta(x_2 - x_3) \, \delta(x - x_2) \>.
\end{equation}
The densities $\rho_0^{i}[\phi_0,J](x)$ are given by the equation,
\begin{equation}\label{ea.e:rhoidef}
   \rho_0^{i}[\phi_0,J](x)
   =
   \frac{1}{2} \,
   \psi_{0\,a}[\phi_0,J](x) \, 
   \calV^i{}^a{}_b \, 
   \psi_0^b[\phi_0,J](x) \>,
\end{equation}
with Grassmann fields $\psi_0^a[j,\phi_0](x)$ given by,
\begin{equation}\label{ea.e:barphidef}
   \psi_0^a[\phi_0,J](x)
   =
   \int [\rd x'] \,
   \calG^a{}_b[\phi_0](x,x') \, j^b(x') \>.
\end{equation}
The densities $\rho_0^{i}[\phi_0,J](x)$ and fields $\psi_0^a[\phi_0,J](x)$ are functionals of both $\phi_0^i(x)$ and all the currents $J^{\alpha}(x)$.  We define the fluctuation inverse Green function $6 \times 6$ matrix $\calD^{-1}_{ij}[\phi_0](x,x')$ by the second-order derivatives evaluated at the stationary points,
\begin{align}
   \calD^{-1}_{ij}[\phi_0](x,x')
   &=
   \frac{ \delta^2 \, \calS_{\text{eff}}[\phi] }
        { \delta \phi^i(x) \, \delta \phi^j(x') } \, \bigg |_{\phi_0} 
   \label{ea.e:calDdef} \\
   &=
   -
   \frac{ g_{ij} }{ \lambda_0 } \, \delta(x,x')
   +
   \Pi_{ij}[\phi_0](x,x') \>,
   \notag
\end{align}
where the polarization tensor $\Pi^{ij}[\phi_0](x,x')$ is given by
\begin{equation}\label{ea.e:Sigmadef}
   \Pi^{ij}[\phi_0](x,x')
   =
   \frac{1}{2} \, 
   \Trb{ \calV^{i} \, \calG[\phi_0](x,x') \,
        \calV^{j} \, \calG[\phi_0](x',x) } \>.
\end{equation}
So inserting these results into the path integral \eqref{ea.e:Z-II} and integrating over the $\phi^i$ fields, the grand potential function is given by
\begin{align}
   &\beta \, \Omega[J]
   \label{ea.e:Omega} \\
   &=
   \beta \, \Omega_0
   +
   \calS_{\text{eff}}[\phi_0,j]
   -
   \frac{1}{2}
   \Tr{ \Ln{ \calD^{-1}[\phi_0](x,x) } }
   +
   \dotsb
   \notag \\
   &=
   \beta \, \Omega_0
   - 
   \frac{1}{2}
   \iint  [\rd x ]\, [ \rd x' ] \,
   j_a(x) \, g^{a}{}_b \, \calG^b{}_c[\phi_0](x,x') \, j^c(x')
   \notag \\
   &
   +
   \int [\rd x] \,
   \Bigl [ \, 
      -
      \frac{ (\phi_i(x) + \mu_i) \,
             g^i{}_j \, 
             (\phi^j(x) + \mu^j) }
           { 2 \, \lambda_0 }
      +
      \phi_{0\,i}(x) S^{i}(x) \,
   \Bigr ]
   \notag \\
   & 
   - 
   \frac{1}{2} \,
   \mathrm{Tr}
   \bigl [ \,
      \Ln{  \calG^{-1}[\phi_0] } \,
   \bigr ] 
   +
   \frac{1}{2}
   \Tr{ \Ln{ \calD^{-1}[\phi_0](x,x) } }
   +
   \dotsb
   \notag
\end{align}
where $\Omega_0$ is an integration constant to be determined.
Instead of writing the thermodynamic potential in terms of the currents $J^{\alpha}(x)$, we can write them in terms of fields $\Psi^{\alpha}(x)$ by Legendre transforming to the thermal vertex potential $\Gamma[\, \Psi \,]$,
\begin{align}
   &\beta \, \Gamma[\, \Psi \,]
   \label{ea.e:GammaI} \\
   &=
   \beta \, \Omega[\, J \,]
   -
   \int [\rd x] \,
   \bigl \{ \,
      \psi_a(x) \, g^a{}_b \, j^b(x)
      +
      \phi_i(x) \, S^i(x) \,
   \bigr \}
   \notag \\
   &
   =
   \beta \, \Omega_0
   +
   \frac{1}{2} 
   \iint [\rd x] \, [\rd x'] \,
   \psi_a(x) \,
   \calG^{-1}[\phi]{}^a{}_b(x,x') \,
   \psi^b(x')
   \notag \\
   & \qquad
   -
   \int [\rd x] \,
      \frac{ (\phi_i(x) + \mu_i) \,
             g^i{}_j \, 
             (\phi^j(x) + \mu^j) }
           { 2 \, \lambda_0 }
   \notag \\
   & \qquad
   -
   \frac{1}{2} \,
   \mathrm{Tr}
   \bigl [ \,
      \Ln{  \calG^{-1}[\phi] } \,
   \bigr ]
   +
   \frac{1}{2} \,
   \mathrm{Tr}
   \bigl [ \,
      \Ln{  \calD^{-1}[\phi] } \,
   \bigr ] \,
   +
   \dotsb \>,
   \notag
\end{align}
which is the classical action plus the trace-log terms.  Currents are given by functional derivatives of $\Gamma[\, \psi, \phi \,]$ with respect to the fields,
\begin{equation}\label{aux.e:dGammadphis}
   \frac{\delta \Gamma[\, \Psi \,]}{\delta \psi_a(x)}
   =
   - g^a{}_b \, j^b(x) \>,
   \quad
   \frac{\delta \Gamma[\, \Psi \,]}{\delta \phi_i(x)}
   =
   - S^i(x) \>.   
\end{equation}
So the derivatives of $\Gamma[\, \Psi \,]$ with respect to the fields vanish for zero currents.  As shown in Ref.~\onlinecite{r:Bender:1977bh}, the last term in Eq.~\eqref{ea.e:GammaI} is of second order in a loop expansion of the effective action in terms of $\phi$-propagators and will be ignored here.  For the original path integral of  Eq.  \eqref{pf.e:Z-I}, the loop expansion in terms of $\psi$ propagators is obtained by realizing that $S$ is measured in units of $\hbar$ and one can evaluate the path integral  as $\hbar \rightarrow  0$ by saddle point (or Laplace's method).  The expansion in $\hbar$ leads to the loop expansion. Similarly one can insert an artificial small parameter $\epsilon$ into the
effective action by replacing $S_\mathrm{eff}$ by $S_\mathrm{eff}/\epsilon$ in Eq.~\eqref{ea.e:Z-II}.   Powers of $\epsilon$ then counts powers of loops in the composite field $\phi$ propagators.  After organizing the series in $\epsilon$ to some specified order one then sets $\epsilon=1$.

%
%%%%%%%%%%%%%%%%%%%%%%%%%%%%%%%%%%%%%%%%%%%%%%%%%%%%%%%%%%%%%%%%%%%%%%
%
\section{\label{s:thermalequib}Uniform system in thermal equilibrium}

For the case of a uniform sample and thermal equilibrium, the fields $\psi^a$ and $\phi^i$ are independent of $x \equiv (\br,\tau)$.  In addition since the Green functions are periodic or anti-periodic in $\tau$, we can expand them in a Fourier series,
\begin{equation} \label{te.e:Gexpand}
   \calG[\phi](x,x')
   =
   \frac{1}{\beta}
   \sum_{\bk,n}
   \tilde{\calG}[\phi](\bk,n) \,
   e^{i [ \bk \cdot ( \br - \br' ) - \omega_n ( \tau - \tau' ) ] } \>,
\end{equation}
where $\omega_n = ( 2n + 1 ) / \beta$ are the Fermi Matsubara frequencies.  So using $\int [\rd x] = \beta \, V$, at thermal equilibrium and for uniform systems, the thermal effective potential from Eq.~\eqref{ea.e:GammaI} is given by
\begin{align}
   &\Veff[\, \Psi \,]
   \equiv
   \Gamma[\, \Psi \,] / V
   =
   \Veffzero
   +
   \frac{1}{2} \,
   \psi_a \,
   \calV[\phi]{}^a{}_b \,
   \psi^b
   \label{te.e:Veff-I} \\
   & \qquad
   -
   \frac{ (\phi_i + \mu_i) \, g^i{}_j \, (\phi^j + \mu^j) }
           { 2 \, \lambda_0 }
   \notag \\
   & \qquad
   -
   \frac{1}{2\beta}
   \Intk \,
   \sum_{n}
   \Tr{ \Ln{ \tilde{\calG}^{-1}[\phi](\bk,n) } }
   +
   \dotsb 
   \notag
\end{align}
Here the matrix $\tilde{\calG}^{-1}[\phi](\bk,n)$ is given by
\begin{equation}\label{te.e:calGinv}
   \tilde{\calG}^{-1}[\phi](\bk,n)
   =
   \begin{pmatrix}
      A(\bk,n) & B(\bk,n) \\
      -B^{\ast}(\bk,n) & - A^{\ast}(\bk,n)
   \end{pmatrix}
\end{equation}
with
\begin{subequations}\label{te.e:AB}
\begin{align}
   A(\bk,n)
   &=
   \begin{pmatrix}
      \epsilon_k + \chi'_{-} - i \omega_n & 0 \\
      0 & \epsilon_k + \chi'_{+} - i \omega_n
   \end{pmatrix} \>,
   \label{te.e:AI} \\
   - A^{\ast}(\bk,n)
   &=
   \begin{pmatrix}
      -\epsilon_k - \chi'_{-} - i \omega_n & 0 \\
      0 & - \epsilon_k - \chi'_{+} - i \omega_n
   \end{pmatrix} \>,
   \label{te.e:AII} \\ 
   B(\bk,n)
   &=
   \begin{pmatrix}
      0 & -\Delta' \\
      \Delta' & 0
   \end{pmatrix} \>,
   \label{te.e:BI} \\
   - B^{\ast}(\bk,n)
   &=
   \begin{pmatrix}
      0 & \Delta^{\prime\,\ast} \\
      -\Delta^{\prime\,\ast} & 0
   \end{pmatrix} \>.
   \label{te.e:BII}   
\end{align}
\end{subequations}
Note that the matrices $A$ and $B$ satisfy the integration conditions for the Grassmann integral.  
From \eqref{AF.e:calVdef}, the matrix $\calV[\phi]$ is given by
\begin{equation}\label{te.e:Vabdef}
   \calV[\phi]
   =
   \begin{pmatrix}
      \chi'_{-} & 0 & 0 & - \Delta' \\
      0 & \chi'_{+} & \Delta' & 0 \\
      0 & \Delta^{\prime\,\ast} &  - \chi'_{-} & 0 \\
      -\Delta^{\prime\,\ast} & 0 & 0 & - \chi'_{+} 
   \end{pmatrix} \>.   
\end{equation}
Then we find
\begin{equation}\label{te.e:psiVpsi}
   \frac{1}{2} \,
   \psi_a \,
   \calV[\phi]{}^a{}_b \,
   \psi^b
   =
   \chi'_{-} \rho_{+} 
   + 
   \chi'_{+} \rho_{-}
   +
   \Delta^{\prime\,\ast} \kappa
   +
   \Delta' \kappa^{\ast} \>.
\end{equation}
Also we find
\begin{align}
   &\frac{ (\phi_i + \mu_i) \, g^i{}_j(\theta) \, (\phi^j + \mu^j) }
           { 2 \, \lambda_0 }
   \label{te.e:phigphi} \\
   & \qquad
   =
   \frac{ (\chi'_{+} + \mu_{+})(\chi'_{-} + \mu_{-})}{\lambda_0 \sin^2\theta}
   +
   \frac{ | \Delta' |^2 }{\lambda_0 \cos^2\theta} \>. 
   \notag 
\end{align}
The grand potential per unit volume is the value of the effective potential evaluated at zero current, or when
\begin{equation}\label{te.e:dVeffdPhi}
   \frac{ \partial \, \Veff[\, \Psi \,] }{ \partial \, \Psi_{\alpha} }
   =
   0 \>,
\end{equation}
for all values of $\alpha$.  From Eq.~\eqref{te.e:psiVpsi}, we find
\begin{equation}\label{cII.e:psiWpsi}
   \frac{1}{2} \,
   \psi_a \,
   \calV[\phi]{}^a{}_b \,
   \psi^b
   =
   \chi'_{-} \rho_{+} 
   +
   \chi'_{+} \rho_{-}
   +
   \Delta' \, \kappa^{\ast}
   +
   \Delta^{\prime\,\ast} \kappa \>,
\end{equation}
and from Eq.~\eqref{te.e:phigphi}, we get
\begin{align}
   &\frac{ (\phi_i + \mu_i ) \, g^i{}_j(\theta) \, (\phi^j + \mu^j ) }
        { 2 \lambda_0 }
   \label{cII.e:phiUphi} \\     
   &=
   \frac{ ( \chi'_{+} + \mu_{+} ) ( \chi'_{-} + \mu_{-} ) }
        { \lambda_0 \sin^2\theta }
   +
   \frac{ | \Delta' |^2 }{ \lambda_0 \cos^2\theta } \>.
   \notag
\end{align}
In order to compute $\Tr{ \Ln{ \tilde{\calG}^{-1}[\phi](\bk,n) } }$, it is simpler to interchange rows and columns of the $\calG^{-1}[\phi]$ matrix so as to bring it into block diagonal form.  To do this, we redefine the fields $\psi^a(x)$ in the following way:
\begin{subequations}\label{cII.e:redefphiul}
\begin{align}
   \psi^a(x)
   &\mapsto
      \Set{ 
      \psi^{\phantom\ast}_{+}(x), 
      \psi^{\ast}_{-}(x),
      \psi^{\ast}_{+}(x),
      \psi^{\phantom\ast}_{-}(x)
       } \>,
       \label{cII.e:redefphiu} \\
   \psi_a(x)
   &\mapsto
      \Set{ 
      \psi^{\ast}_{+}(x), 
      \psi^{\phantom\ast}_{-}(x),
      \psi^{\phantom\ast}_{+}(x),
      \psi^{\ast}_{-}(x) 
       } \>.
       \label{cII.e:redefphil}
\end{align}
\end{subequations}
Then from Eqs.~\eqref{te.e:calGinv} and \eqref{te.e:AB}, the Fourier transform of the inverse $\calG$-matrix is of the form,
\begin{equation}\label{cII.e:calGinvdef}
   \tilde{\calG}^{-1}(\bk,n)
   =
   \begin{pmatrix}
      \tilde{G}^{-1}(\bk,n) & 0 \\
      0 &  - \tilde{G}^{-1\,\dagger}(\bk,n)
   \end{pmatrix} \>,
\end{equation}
where
\begin{equation}\label{cII.e:Ginvdef}
   \tilde{G}^{-1}(\bk,n)
   =
   \begin{pmatrix}
      \epsilon_k + \chi'_{-} - i \omega_n & - \Delta' \\
      - \Delta^{\prime\,\ast} & - \epsilon_k - \chi'_{+} - i \omega_n 
   \end{pmatrix} \>.
\end{equation}
After some algebra, we find
\begin{align}
   \Det{ \tilde{G}^{-1}(\bk,n) }
   &=
   - ( \, \alpha_1 + i \alpha_2 \, ) \>,
   \label{cII.e:detGs} \\
   \Det{ \tilde{G}^{-1\,\dagger}(\bk,n) }
   &=
   - ( \, \alpha_1 - i \alpha_2 \, ) \>, 
   \notag  
\end{align}
where
\begin{subequations}\label{cII.e:alphabeta}
\begin{align}
   \alpha_1
   &=
   ( \epsilon_k + \chi'_{+} ) ( \epsilon_k + \chi'_{-} )
   +
   | \Delta' |^2 
   +
   \omega_n^2 \>,
   \label{cII.e:alpha} \\
   \alpha_2
   &=
   ( \chi'_{-} - \chi'_{+} ) \, \omega_n \>.
\end{align}
\end{subequations}
So we find
\begin{align}
   &\Trb{ \Ln{ \tilde{\calG}^{-1}(\bk,n) } }
   =
   \Ln{ \Det{ \tilde{\calG}^{-1}(\bk,n) } }
   =
   \label{cII.e:trlogG} \\
   &\qquad
   =
   \Ln{ 
      \Det{ \tilde{G}^{-1}(\bk,n) } \, 
      \Det{ \tilde{G}^{-1\,\dagger}(\bk,n) } 
      }
   \notag \\
   &\qquad
   =
   \Ln{ \alpha_1^2 + \alpha_2^2 }
   =
   \Ln{ \omega_n^4 + 2 \, b \, \omega_n^2 + c }
   \notag \\
   &\qquad
   =
   \Ln{ 
   ( \omega_n^2 + \omega_{+}^2 ) \,
   ( \omega_n^2 + \omega_{-}^2 ) }
   \notag \\
   &\qquad
   =
   \Ln{ \omega_n^2 + \omega_{+}^2 }
   +
   \Ln{ \omega_n^2 + \omega_{-}^2 } \>,
   \notag 
\end{align}
where $\omega_{\pm}^2$ are the two roots of the equation
\begin{equation}\label{cII.e:tworoots}
   \omega_{\pm}^4
   -
   2 \, b \, \omega_{\pm}^2
   +
   c
   =
   0 \>,
   \Quad{$\Rightarrow$}
   \omega_{\pm}^2
   =
   b
   \pm
   \sqrt{b^2 - c} \>,
\end{equation}
with
\begin{align}
   b
   &=
   ( \epsilon_k + \chi'_{+} ) ( \epsilon_k + \chi'_{-} )
   +
   | \Delta' |^2 
   +
   ( \chi'_{+} - \chi'_{-} )^2 / 2 
   \label{cII.e:abdefs} \\
   &=
   \frac{1}{2} \,
   \bigl [ \,
      ( \epsilon_k + \chi'_{+} )^2
      +
      ( \epsilon_k + \chi'_{-} )^2
      +
      2 \, | \Delta' |^2 \,
   \bigr ]
   \notag \\
   c
   &=
   \bigl [ \, 
      ( \epsilon_k + \chi'_{+} ) ( \epsilon_k + \chi'_{-} )
      +
      | \Delta' |^2 \,
   \bigr ]^2 \>.
   \notag
\end{align}
The square root of the discriminant is given by
\begin{align}
   &\sqrt{b^2 - c}
   \label{cII.e:discriminant} \\
   &=
   \frac{1}{2} \,
   | \chi'_{+} - \chi'_{-} | \,
   \sqrt{
      [ \,
         ( \epsilon_k + \chi'_{+} )
         +
         ( \epsilon_k + \chi'_{-} ) \,
      ]^2
      +
      4 \, | \Delta' |^2 \,
         } \>.
      \notag
\end{align}
so the frequencies $\omega_{\pm}^2$, which depend on $k$, are given by
\begin{align}
   \omega_{\pm}^2
   &=
   \frac{1}{2} \,
   \Bigl \{ \,
   \bigl [ \,
      ( \epsilon_k + \chi'_{+} )^2
      +
      ( \epsilon_k + \chi'_{-} )^2
      +
      2 \, | \Delta' |^2 \,
   \bigr ]
   \label{cII.e:omegapm} \\
   & \!\!
   \pm
   | \chi'_{+} - \chi'_{-} | \,
   \sqrt{
      [ \,
         ( \epsilon_k + \chi'_{+} )
         +
         ( \epsilon_k + \chi'_{-} ) \,
      ]^2
      +
      4 \, | \Delta' |^2 \,
         } \,
   \Bigr \} \>.
   \notag      
\end{align}
So from \eqref{cII.e:trlogG}
\begin{align}
   &\frac{1}{2} \,
   \Tr{ \Ln{ \tilde{\calG}^{-1}(\bk,n) } }
   \label{cII.e:TrLn} \\
   &
   =
   \frac{1}{2\beta} \Intk \, \sum_n 
   \Ln{ \Det{ \tilde{\calG}^{-1}(\bk,n) } }
   \notag \\
   &
   =
   \frac{1}{2\beta} \Intk \, 
   \Bigl \{ \,
      \sum_n 
      \Ln{ \omega_n^2 + \omega_{+}^2 }
      +
      \sum_n 
      \Ln{ \omega_n^2 + \omega_{-}^2 } \,
   \Bigr \}
   \notag \\
   &
   =
   \Intk \,
   \Bigl \{ \,
      \frac{ \omega_{+} + \omega_{-} }{2}
      \notag \\
      & \qquad\qquad
      +
      \frac{1}{\beta} \,
      \bigl \{ \,
         \Ln{ 1 + e^{ -\beta \omega_{+}} }
         +
         \Ln{ 1 + e^{ -\beta \omega_{-}} } \,
      \bigr \} \, 
   \Bigr \} \>.
   \notag
\end{align}
Then from \eqref{te.e:Veff-I}, the effective potential becomes
\begin{align}
   &\Veff[\, \Psi \,]
   =
   \Veffzero
   +
   \chi'_{+} \rho_{-} 
   +
   \chi'_{-} \rho_{+}
   +
   \Delta' \, \kappa^{\ast}
   +
   \Delta^{\prime\,\ast} \kappa
   \label{cII.e:Veff-II} \\
   & \qquad
   -
   \frac{ ( \chi'_{+} + \mu_{+} ) ( \chi'_{-} + \mu_{-} ) }
        { \lambda_0 \sin^2\theta }
   -
   \frac{ | \Delta' |^2 }{ \lambda_0 \cos^2\theta }
   \notag \\
   & \qquad
   -
   \Intk \,
   \Bigl \{ \,
      \frac{ \omega_{+} + \omega_{-} }{2}
      \notag \\
      & \qquad\qquad
      +
      \frac{1}{\beta} \,
      \bigl \{ \,
         \Ln{ 1 + e^{ -\beta \omega_{+}} }
         +
         \Ln{ 1 + e^{ -\beta \omega_{-}} } \,
      \bigr \} \, 
   \Bigr \} \>.
   \notag   
\end{align}
Expanding $\omega_{\pm}$ in a Laurent series about $k \rightarrow \infty$, we find
\begin{equation}\label{cII.e:resultII}
   \frac{\omega_{+} + \omega_{-}}{2}
   =
   \epsilon_k
   +
   \frac{1}{2} \, [ \, \chi'_{+} + \chi'_{-} \, ]
   +
   \frac{  | \Delta' |^2 }
        { 2 \, \epsilon_k }
   +
   \dotsb
\end{equation}
So using dimensional regularization~\cite{r:Papenbrock:1999fk}, the effective potential becomes
\begin{align}
   &\Veff[\, \Psi \,]
   =
   \chi'_{+} \rho_{-} 
   +
   \chi'_{-} \rho_{+}
   +
   \Delta' \, \kappa^{\ast}
   +
   \Delta^{\prime\,\ast} \kappa
   \label{cII.e:Veff-III} \\
   & \quad
   -
   \frac{ ( \chi'_{+} + \mu_{+} ) ( \chi'_{-} + \mu_{-} ) }
        { \lambda \sin^2\theta }
   -
   \frac{ | \Delta' |^2 }{ \lambda \cos^2\theta }
   \notag \\
   & \quad
   -
   \Intk \,
   \Bigl \{ \,
      \frac{ \omega_{+} + \omega_{-} }{2}
      -      
      \epsilon_k
      -
      \frac{1}{2} \, [ \, \chi'_{+} + \chi'_{-} \, ]
      -
      \frac{  | \Delta' |^2 }
           { 2 \, \epsilon_k }
      \notag \\
      & \qquad\qquad
      +
      \frac{1}{\beta} \,
      \bigl \{ \,
         \Ln{ 1 + e^{ -\beta \omega_{+}} }
         +
         \Ln{ 1 + e^{ -\beta \omega_{-}} } \,
      \bigr \} \, 
   \Bigr \} \>,
   \notag   
\end{align}
where the coupling constant is related to the s-wave scattering length, $a$,  i.e. $\lambda = 8\pi \gamma \, a /$.
Recall that $\rho_{\pm} = \psi^{\ast}_{\pm} \, \psi^{\phantom\ast}_{\pm}$ and $\kappa = \psi_{+} \, \psi_{-}$. 

We recover the thermodynamic grand potential by evaluating $\calV_{\text{eff}}[\, \Psi \,]$ at the minimum of the potential when
\begin{equation}\label{cII.e:dVeffdpsidpi}
   \frac{\partial \, \calV_{\text{eff}}[\, \Psi \,]}{\partial \, \psi_{a}}
   =
   0 \>,
   \Qquad{and}
   \frac{\partial \,  \calV_{\text{eff}}[\, \Psi \,]}{\partial \, \phi_{i}}
   =
   0 \>,   
\end{equation}
for all values of $a$ and $i$.  For the Grassmann $\psi_a$ fields, derivatives of the effective potential \eqref{cII.e:Veff-III} with respect to $\psi_{\pm}$ give
\begin{equation}\label{cII.e:dVdpsi}
   \bigl [ \,
      \chi'_{+} \chi'_{-}
      +
      | \Delta' |^2 \,
   \bigr ] \, \psi_{\pm}
   =
   0 \>.
\end{equation}
The above can be satisfied only if $\psi_{\pm} = 0$.

\section{\label{ss:caseA}Equal chemical potentials}

In this section we set $\mu_{+} = \mu_{-} \equiv \mu$ and $\chi'_{+} = \chi'_{-} \equiv \chi'$, so that only the total particle density is fixed.  For this case the only possible solution for the Grassmann fields is the first case above where $\psi_{\pm} = 0$.  The frequency spectrum is given by
\begin{equation}\label{cA.e:omega}
   \omega^2_k
   \equiv
   \omega_{+}^2
   =
   \omega_{-}^2
   =
   ( \epsilon_k + \chi' )^2
   +
   | \Delta' |^2 \>,
\end{equation}
and from \eqref{cII.e:Veff-III}, the effective potential for this case is given by
\begin{align}
   \Veff[\, \Psi \,]
   &=
   -
   \frac{ ( \chi' + \mu )^2 }
        { \lambda \sin^2\theta }
   -
   \frac{ | \Delta' |^2 }{ \lambda \cos^2\theta }
   \label{cIIA.e:Veff-III} \\
   &
   -
   2 \Intk \,
   \Bigl \{ \,
      \frac{1}{2} \,
      \Bigl [ \,
         \omega_{k}
         -      
         \epsilon_k
         -
         \chi' 
         -
         \frac{  | \Delta' |^2 }
              { 2 \, \epsilon_k } \,
      \Bigr ]
      \notag \\
      & \qquad
      +
      \frac{1}{\beta} \,
         \Ln{ 1 + e^{ -\beta \omega_k } }
   \Bigr \} \>.
   \notag   
\end{align}
The gap equation for the $\chi'$ field is now,
\begin{align}
   &\frac{ \chi' + \mu }{ \lambda \sin^2\theta }
   =
   \frac{1}{2}
   \Intk \,
   \Bigl \{ \,
      \Bigl ( \frac{ \partial \, \omega_k }{ \partial \chi' } \Bigr ) \,
      \bigl [ \, 2 n(\beta \omega_k) - 1 \, \bigr ] 
      +
      1 \,
   \Bigr \}
   \notag \\
   &=
   \frac{1}{2}
   \Intk \,
   \Bigl \{ \,
      \frac{ \epsilon_k + \chi' }{ \omega_k } \,
      \bigl [ \, 2 n(\beta \omega_k) - 1 \, \bigr ] 
      +
      1 \,
   \Bigr \} \>,
   \label{cIIA.e:chigapI}
\end{align}
where the non-interacting Fermi particle number factor $n(x)$ is defined by
\begin{equation}\label{cIIA.e:nFermi}
   n(x)
   =
   1 / [ \, e^{x} + 1 \, ] \>.
\end{equation}
The gap equation for $\Delta'$ is 
\begin{align}
   &\frac{ \Delta' }{ \lambda \cos^2\theta }
   =
   \Intk \,
   \Bigl \{ \,
      \Bigl ( \frac{ \partial \, \omega_k }{ \partial \Delta^{\prime\,\ast} } \Bigr ) \,
      \bigl [ \, 2 n(\beta \omega_k) - 1 \, \bigr ] 
      +
      \frac{\Delta'}{2 \epsilon_k} \,
   \Bigr \}
   \notag \\
   &=
   \Intk \,
   \Bigl \{ \,
      \frac{ \Delta' }{2 \omega_k} \,
      \bigl [ \, 2 n(\beta \omega_k) - 1 \, \bigr ] 
      +
      \frac{\Delta'}{2 \epsilon_k} \,
   \Bigr \} \>.
   \label{cIIA.e:DeltagapI}
\end{align}
Again, the factor of $\Delta'$ cancels, and we get for the $\Delta'$ gap equation
\begin{equation}\label{cIIA.e:DeltagapII}
   \frac{1}{ \lambda \cos^2\theta }
   =
   \frac{1}{2}
   \Intk \,
   \Bigl \{ \,
      \frac{1}{\omega_k} \,
      \bigl [ \, 2 n(\beta \omega_k) - 1 \, \bigr ] 
      +
      \frac{1}{\epsilon_k} \,
   \Bigr \} \>.   
\end{equation}
The total particle density is given by
\begin{align}
   \rho
   &=
   - \frac{\partial \, \Veff[ \Psi ]}{\partial \mu}
   =
   2 \, \frac{\chi' + \mu}{\lambda \sin^2\theta}
   \label{cIIA.e:rho} \\
   &
   =
   \Intk \,
   \Bigl \{ \,
      \frac{ \epsilon_k + \chi' }{ \omega_k } \,
      \bigl [ \, 2 n(\beta \omega_k) - 1 \, \bigr ] 
      +
      1 \,
   \Bigr \} \>.
   \notag
\end{align}
It is convenient to scale momenta and energies in terms of the Fermi momentum, $\kF$, and Fermi energy, $\epsilonF = \gamma \kF^2$, respectively. We introduce
\begin{gather}
   \bar{k} = k / \kF \>,
   \quad
   \bar{\mu}
   =
   \mu / \epsilonF \>,
   \quad
   \bar{\Delta} 
   =
   \Delta / \epsilonF \>,
   \label{scale.e:scaling} \\
   \bar{\omega}_{\bar{k}}
   =
   \omega_{k} / \epsilonF
   =
   \sqrt{( \bar{k}^2 + \bar{\chi}' )^2+ \bar{\Delta}^2 } \>,
   \notag \\
   \bar{T} = T / \epsilonF = T / T_\mathrm{F}  \>,
   \quad
   \bar{\beta} = \epsilonF \, \beta = T_\mathrm{F}  / T = 1 / \bar{T} \>.
   \notag
\end{gather}
Then, the rescaled equations are
\begin{subequations}\label{cIIA.e:gapdenI}
\begin{gather}
   \frac{1}{\xi \, \cos^2\theta}
   =
   \frac{2}{\pi}
   \int_{0}^{\infty} \!\! \bar{k}^2 \, \rd \bar{k} \,
   \Bigl \{ \,
      \frac{1}{ \bar{k}^2 }
      -
      \frac{ 1 - 2 \, n( \bar{\beta} \, \bar{\omega}_{\bar{k}} ) }
           { \bar{\omega}_{\bar{k}} } \,
   \Bigr \} \>,
   \label{cIIA.e:gapA} \\
   1
   =
   \frac{3}{2}
   \int_{0}^{\infty} \!\! \bar{k}^2 \, \rd \bar{k} \,
   \Bigl \{ \,
      1
      -
      \frac{ \bar{k}^2 + \bar{\chi}' }{ \bar{\omega}_{\bar{k}} } \,
      [ \, 1 - 2 \, n( \bar{\beta} \, \bar{\omega}_{\bar{k}} ) \, ]
   \Bigr \} \>,
   \label{cIIA.e:gapB} \\
   \bar{\chi}'
   =
   \frac{4}{3\pi} \, \xi \, \sin^2\theta
   -
   \bar{\mu} \>,
   \label{cIIA.e:gapC} 
\end{gather} 
\end{subequations} 
where now $\bar{\omega}_{\bar{k}}^2 = ( \bar{k}^2 + \bar{\chi}' )^2 + |\bar{\Delta}|^2$ and we introduced $\xi = k_\mathrm{F} a$.  Eqs.~\eqref{cIIA.e:gapdenI} are to be solved selfconsistently for $\bar{\mu}$ and $|\bar{\Delta}'|$.  

%
%%%%%%%%%%%%%%%%%%%%%%%%%%%%%%%%%%%%%%%%%%%%%%%%%%%%%%%%%%%%%%%%%%%%%%
%
% Fig 1
%
\begin{figure}[t!]
   \centering
   \includegraphics[width=0.95\columnwidth]{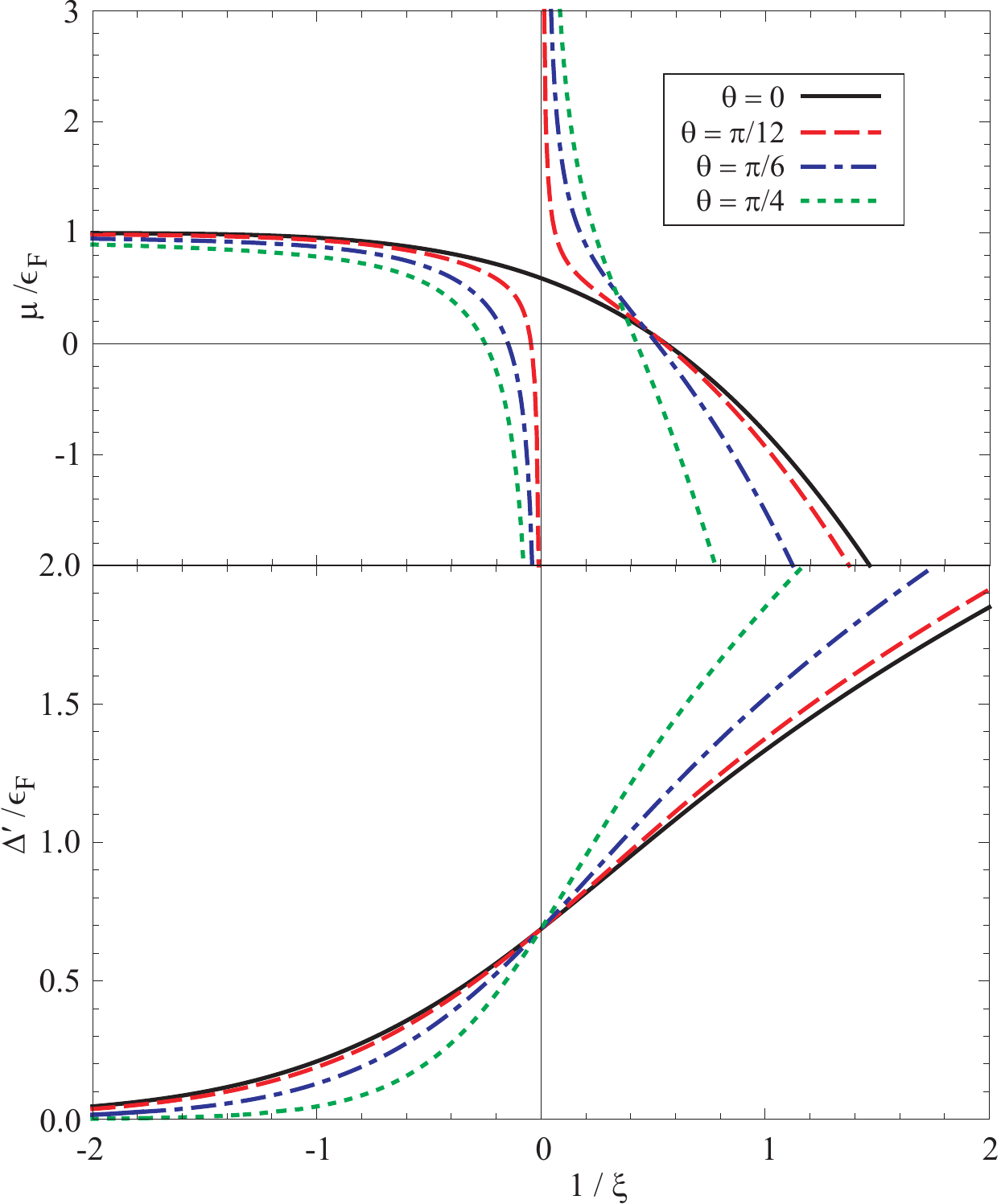}
   \caption{\label{f:T0deltamu-eta}(Color online) Zero temperature solutions for $\Delta'$ 
   and $\mu$ of the gap equations in scaled  units 
   \vs\ $1/\xi$ for several values of the parameter $\theta = 0$.} 
\end{figure}
%
%%%%%%%%%%%%%%%%%%%%%%%%%%%%%%%%%%%%%%%%%%%%%%%%%%%%%%%%%%%%%%%%%%%%%%
%
% Fig 2
%
\begin{figure}[t!]
   \centering
   \includegraphics[width=0.95\columnwidth]{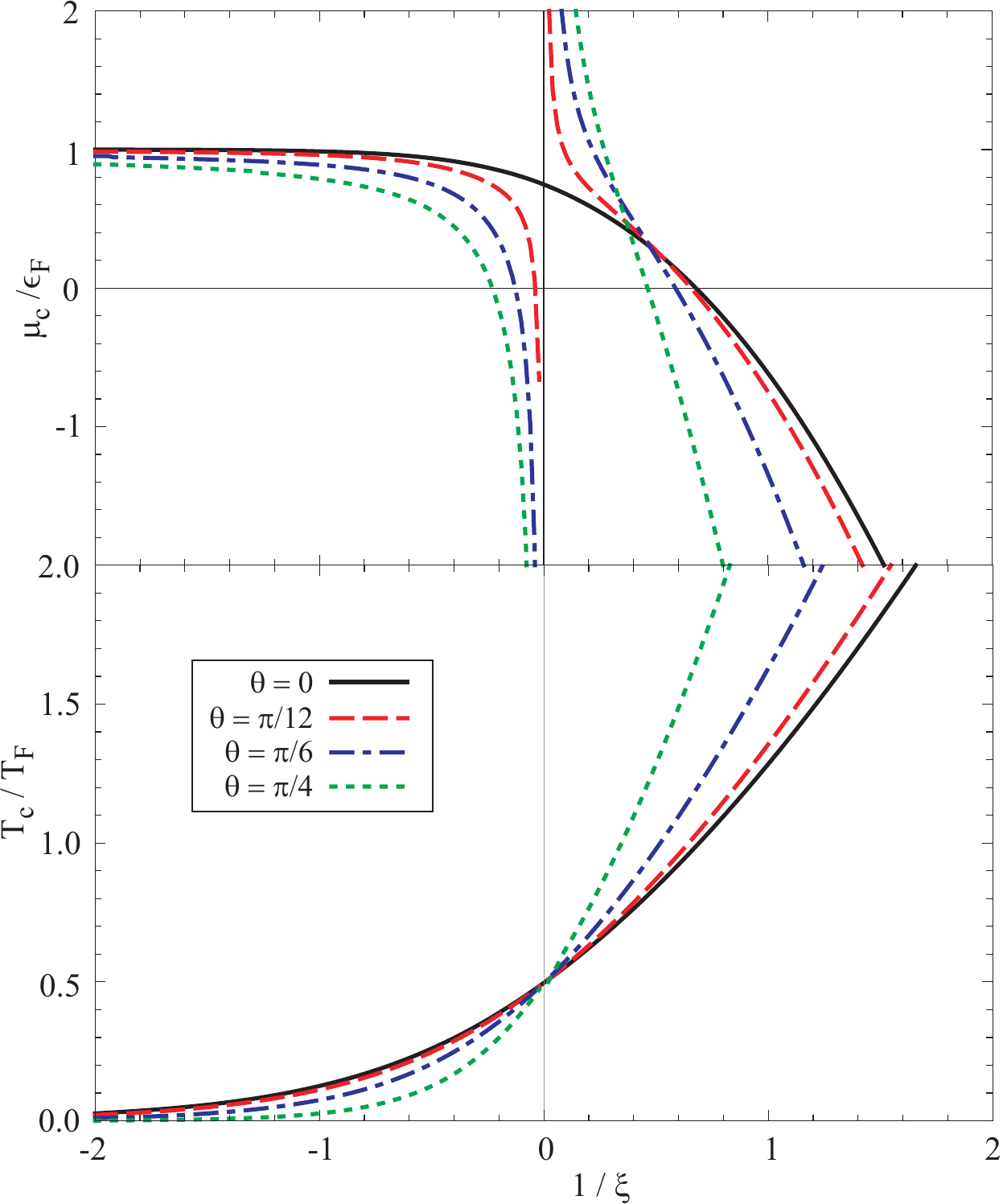}
   \caption{\label{f:Deltazero}(Color online) Solutions for $T$ and $\mu$ of the gap equations 
   in scaled units at the critical point ($\Delta'=0$)
   \vs\ $1/\xi$ for several values of the parameter $\theta = 0$.} 
\end{figure}

%
%%%%%%%%%%%%%%%%%%%%%%%%%%%%%%%%%%%%%%%%%%%%%%%%%%%%%%%%%%%%%%%%%%%%%%
%
\subsection{\label{ss:T0}Zero temperature ($T = 0$)}

At zero temperature so that $n( \bar{\beta} \, \bar{\omega}_k ) = 0$, Eqs.~\eqref{cIIA.e:gapdenI} reduce to
\begin{align}
   \frac{1}{\xi \, \cos^2\theta}
   &=
   \frac{2}{\pi}
   \int_{0}^{\infty} \!\! \rd \bar{k} \,
   \Bigl \{ \,
      1
      -
      \frac{\bar{k}^2}{ \bar{\omega}_{\bar{k}} } \,
   \Bigr \} \>,
   \label{cIIA.e:gapdenzerotemp} \\
   1
   &=
   \frac{3}{2}
   \int_{0}^{\infty} \!\! \bar{k}^2 \, \rd \bar{k} \,
   \Bigl \{ \,
      1
      -
      \frac{ \bar{k}^2 + \bar{\chi}' }{ \bar{\omega}_{\bar{k}} } \,
   \Bigr \} \>,
   \notag \\
   \bar{\chi}'
   &=
   \frac{4}{3\pi} \, \xi \, \sin^2\theta
   -
   \bar{\mu} \>.
   \notag
\end{align}  
In Fig.~\ref{f:T0deltamu-eta} we illustrate the solutions of the gap equations \eqref{cIIA.e:gapdenzerotemp} for $\Delta'$ and $\mu$ in reduced units as a function of $1/\xi$ for values of $\theta = 0$, $\pi/12$, $\pi/6$ and $\pi/4$.  
For $\theta = 0$, our results reduce to the variational equations discussed by Leggett~\cite{r:Leggett:1980fk}. In the unitarity limit (i.e. for $1/\xi = 0$) one obtains $\mu / \epsilon_F = 0.59$ and $\Delta / \epsilon_F = 0.69$.  

For $\theta \ne 0$, the chemical potential has a singularity at $1/\xi = 0$. This indicates that the only physical theory corresponds to choosing $\theta = 0$. Hence, the BCS theory is the only relevant auxiliary field theory for a dilute gas of fermions.

%
%%%%%%%%%%%%%%%%%%%%%%%%%%%%%%%%%%%%%%%%%%%%%%%%%%%%%%%%%%%%%%%%%%%%%%
%
\subsection{\label{ss:Tc}Critical temperature ($\Delta' = 0$)}

At finite temperature, the critical temperature and critical chemical potential correspond to the point where the gap $\Delta = 0$. At the critical point, the spectrum becomes $\omega_k = \epsilon_k + \chi'$.  For this case, Eqs.~\eqref{cIIA.e:gapdenI} become
\begin{gather}
   \frac{1}{\xi \, \cos^2\theta}
   =
   \frac{2}{\pi}
   \int_{0}^{\infty} \!\! k^2 \, \rd k \,
   \Bigl \{ \,
      \frac{1}{ k^2 }
      -
      \frac{ \tanh( \beta \, \omega_k / 2 ) }
           { \omega_k } \,
   \Bigr \}
   \label{scale.e:zeroDelta} \\
   1
   =
   \frac{3}{2}
   \int_{0}^{\infty} \!\! \bar{k}^2 \, \rd \bar{k} \,
   \bigl \{ \,
      1 - \Sgn{\bar{k}^2 + \bar{\chi}'} \, \tanh( \beta \, \bar{\omega}_{\bar{k}} / 2 ) \,
   \bigr \} \>,
   \notag \\
   \bar{\chi}'
   =
   \frac{4}{3\pi} \, \xi \, \sin^2\theta
   -
   \bar{\mu} \>.
   \notag
\end{gather}  
Solutions of Eqs.~\eqref{scale.e:zeroDelta} for various values of $\theta$ are shown in Fig.~\ref{f:Deltazero}.  For $\theta = 0$, our results are the same as those discussed extensively by S\'a de Melo, Randeria and Engelbrecht in Refs.~\onlinecite{r:Melo:1993vn,r:Engelbrecht:1997fk}.  For $\theta \neq 0$, the chemical potential at the critical temperature has a singularity in the unitarity limit, again indicating that the case of $\theta=0$ corresponds to the only physical theory for a dilute gas of fermions in the auxiliary field formalism.

%
%%%%%%%%%%%%%%%%%%%%%%%%%%%%%%%%%%%%%%%%%%%%%%%%%%%%%%%%%%%%%%%%%%%%%%
%
\subsection{\label{ss:EOS}Thermodynamics}

\begin{figure}[t!]
   \centering
   \includegraphics[width=0.95\columnwidth]{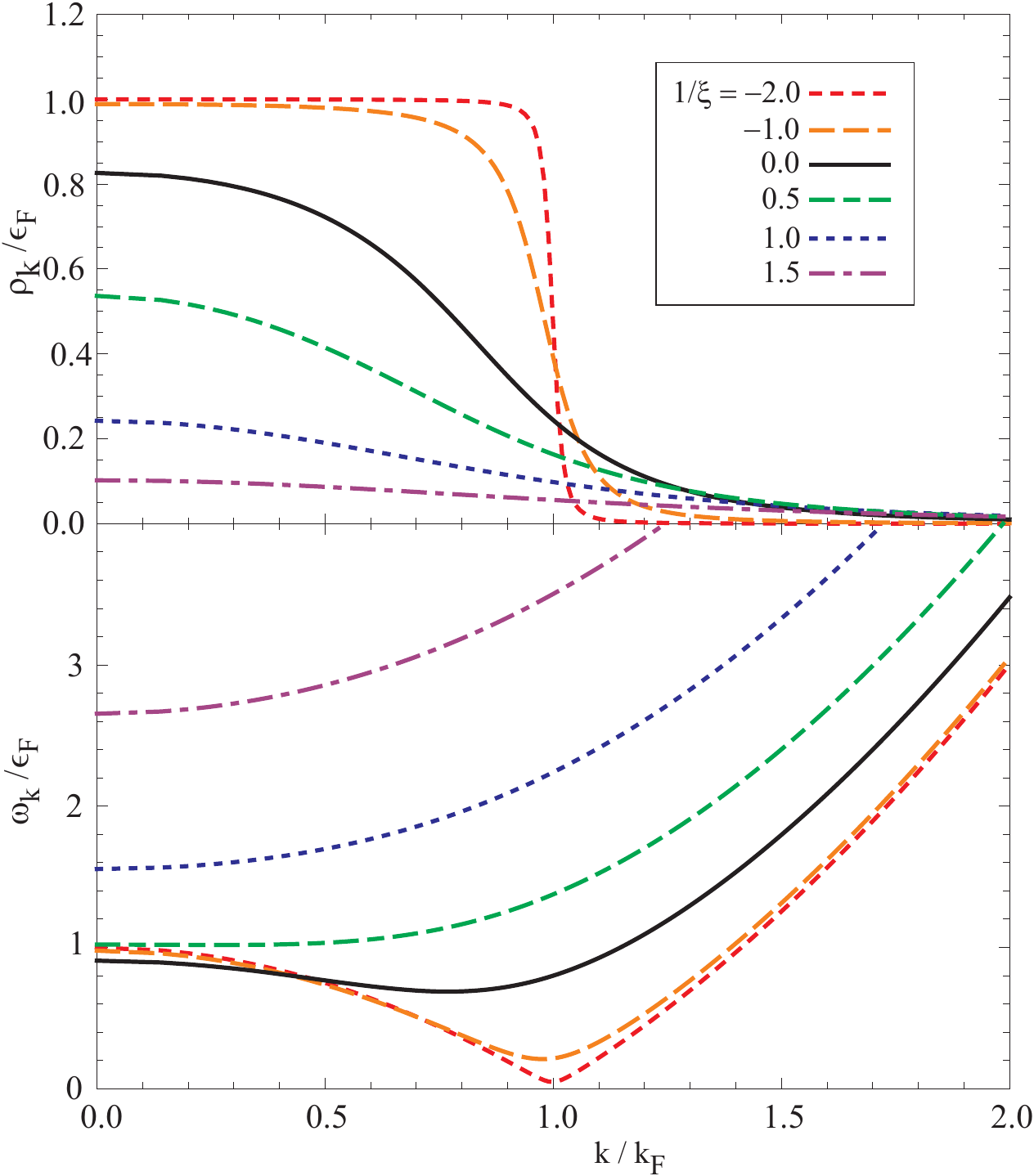}
   \caption{\label{f:T0theta0-rhok}(Color online)  
   Momentum dependence of the zero temperature particle distributions, $\rho(k)$, and dispersion relations, $\omega_k$, for $\theta=0$ and  
   several values of the parameter~$1/\xi$. We note that the location of the minimum in the dispersion relation shifts smoothly to zero momentum 
   and disappears for $\xi > \xi_c \approx$0.55, indicative of the crossover character of the BCS to BEC transition.}
\end{figure}

From Eq.~\eqref{cIIA.e:Veff-III}, the particle number density is
\begin{align}
   \rho
   &=
   \frac{N}{V}
   =
   - \frac{1}{V} \Partial{\,\Omega}{\mu}{T}{V}
   =
   - \frac{\partial \Veff}{\partial \mu}
   =
   2 \, \frac{\chi' + \mu}{\lambda \, \sin^2\theta}
   \label{XXX.e:N} \\
   &=
   2 \Intk \, \rho(k) \>,
   \notag 
\end{align}
where
\begin{equation}\label{XXX.e:rhokdef}
   \rho(k)
   =
   \frac{1}{2} \,
   \Bigl \{ \, 
      1
      -
      \frac{ \epsilon_k + \chi' }{ \omega_k } \,
      \bigl [ \, 1 - 2 n(\beta \omega_k) \, \bigr ] \,
   \Bigr \} \>.   
\end{equation}
The zero-temperature momentum distribution of the particle distribution function, $\rho(k)$, is shown in Fig.~\ref{f:T0theta0-rhok} for several values of the parameter~$\xi$. For completeness, we also depict the momentum dependence of the dispersion relations, $\omega_k$, for $\theta=0$ and  the same values of the parameter~$\xi$. We note that the location of the minimum in the dispersion relation shifts smoothly to zero momentum  and disappears for $\xi > \xi_c \approx$0.55, indicative of the crossover character of the BCS to BEC transition~\cite{r:Parish:2005fk}. 

The pressure is also obtained from Eq.~\eqref{cIIA.e:Veff-III}, as
\begin{align}
   p
   &=
   - \Partial{\,\Omega}{V}{T}{\mu}
   =
   - \Veff
   =
   \frac{ ( \chi' + \mu )^2 }
        { \lambda \sin^2\theta }
   +
   \frac{ | \Delta' |^2 }{ \lambda \cos^2\theta }
   \label{XXX.e:pres}  \\
   & \qquad
   +
   2 \Intk \,
   \Bigl \{ \,
      \frac{1}{2} \,
      \Bigl [ \,
         \omega_{k}
         -      
         \epsilon_k
         -
         \chi' 
         -
         \frac{  | \Delta' |^2 }
              { 2 \, \epsilon_k } \,
      \Bigr ]
      \notag \\
      & \qquad\qquad\qquad\qquad\qquad
      +
      \frac{1}{\beta} \,
         \Ln{ 1 + e^{ -\beta \omega_k } }
   \Bigr \} \>,
   \notag
\end{align}
In scaled variables, the pressure is given by
\begin{align}
   \frac{p}{\rho \, \epsilonF}
   &=
   \frac{2}{3 \pi} \, \xi \, \sin^2\theta
   +
   \frac{3 \pi}{8 \, \xi} \,
   \frac{ | \bar{\Delta}' |^2 }{ \cos^2\theta } 
   \label{XXX.e:pressII} \\
   & \quad
   +
   \frac{3}{2}
   \int_{0}^{\infty} \!\! \bar{k}^2 \, \rd \bar{k} \,
   \Bigl \{ \,
      \bar{\omega}_{\bar{k}}
      -      
      \bar{k}^2
      -
      \bar{\chi}' 
      -
      \frac{ | \bar{\Delta}' |^2 }{ 2 \, \bar{k}^2 } \,
      \notag \\
      & \qquad\qquad
      +
      \frac{2}{\bar{\beta}} \,
         \Ln{ 1 + e^{ -\bar{\beta} \, \bar{\omega}_{\bar{k}} } }  \,
   \Bigr \} \>,
   \notag    
\end{align}
From Eq.~\eqref{thm.e:SNpI}, the entropy per unit volume, $s$, is given by
\begin{align}
   s
   &=
   \frac{S}{V}
   =
   - \frac{1}{V} \, \Partial{\,\Omega}{T}{\mu}{V}
   =
   \frac{\beta^2}{V} \, \frac{\partial \, \Omega}{\partial \beta}
   =
   \beta^2 \, \frac{\partial \, \Veff}{\partial \beta}
   \notag \\
   &=
   2 \beta
   \Intk \,
   \Bigl \{ \,
      n(\beta \omega_k) \, \omega_k
      +
      \frac{1}{\beta} \,
      \Ln{ 1 + e^{-\beta \omega_k} } \,
   \Bigr \}
   \notag \\
   &=
   - 2
   \Intk \,
   \bigl \{ \,
      n(\beta \omega_k) \, \Ln{ n(\beta \omega_k) }
      \label{XXX.e:S} \\
      & \qquad\qquad
      +
      [ \, 1 - n(\beta \omega_k) \, ] \,
      \Ln{ 1 - n(\beta \omega_k) } \,
   \bigr \} \>,
   \notag   
\end{align}
or, in scaled units, 
\begin{equation}
   \frac{s}{\rho \, \epsilonF}
   =
   \frac{3}{T}
   \int_{0}^{\infty} \!\! \bar{k}^2 \, \rd \bar{k} \,
   \Bigl \{ \,
      n( \, \bar{\beta}\bar{\omega}_{\bar{k}} \, ) \, 
      \bar{\omega}_{\bar{k}}
      +
      \frac{1}{\bar{\beta}} \,
      \Ln{ 1 + e^{-\bar{\beta} \bar{\omega}_{\bar{k}} } } \,
   \Bigr \} \>,
\end{equation}
From Eq.~\eqref{thm.e:U}, the energy per unit volume, $e$, is given by
\begin{align}
   e
   &=
   E/V 
   =
   \Veff + T \, s  + \mu \, \rho 
   \label{XXX.e:u} \\
   &=
   -
   \frac{ \chi^{\prime\,2} - \mu^2 }
        { \lambda \sin^2\theta }
   -
   \frac{ | \Delta' |^2 }{ \lambda \cos^2\theta }
   \notag \\
   &
   +
   \Intk \,
   \Bigl \{ \,
      [ \, 2 n(\beta \omega_k) - 1 \, ] \, \omega_k
      +      
      \epsilon_k
      +
      \chi' 
      +
      \frac{  | \Delta' |^2 }
           { 2 \, \epsilon_k } \,
   \Bigr \} \>,
   \notag 
\end{align}
or, in scaled units, 
\begin{align}
   &\frac{e}{\rho \, \epsilonF}
   =
   -
   \frac{p}{\rho \, \epsilonF}
   +
   T \, \frac{s}{\rho \, \epsilonF}
   +
   \frac{\mu}{\epsilonF} 
   \label{XXX.e:uIII} \\
   & \qquad 
   =
   -
   \frac{2}{3\pi} \, \xi \, \sin^2\theta
   +
   \bar{\mu}
   -
   \frac{3\pi}{8 \, \xi} \,
   \frac{ | \bar{\Delta}' |^2 }{ \cos^2\theta }
   \notag \\
   &
   -
   \frac{3}{2}
   \int_{0}^{\infty} \!\! \bar{k}^2 \, \rd \bar{k} \,
   \Bigl \{ \,
      \bar{\omega}_{\bar{k}} \,
      [ \, 1 - 2 n(\bar{\beta}\bar{\omega}_{\bar{k}}) \, ] \,
      -      
      \bar{k}^2
      -
      \bar{\chi}' 
      -
      \frac{ | \bar{\Delta}' |^2 }{ 2 \, \bar{k}^2 } \,
   \Bigr \} \>.
   \notag 
\end{align}
Here $\bar{\chi}'$ and $\bar{\Delta}'$ solutions of Eqs.~\eqref{cIIA.e:gapdenI}.  
Comparing Eqs.~\eqref{XXX.e:pressII} and \eqref{XXX.e:uIII}, we see that at $T=0$, 
\begin{equation}\label{XXX.e:Tzero-u-p}
   e
   =
   - p + \mu \, \rho \>.
\end{equation}

\begin{figure}[t!]
   \centering
   \includegraphics[width=0.95\columnwidth]{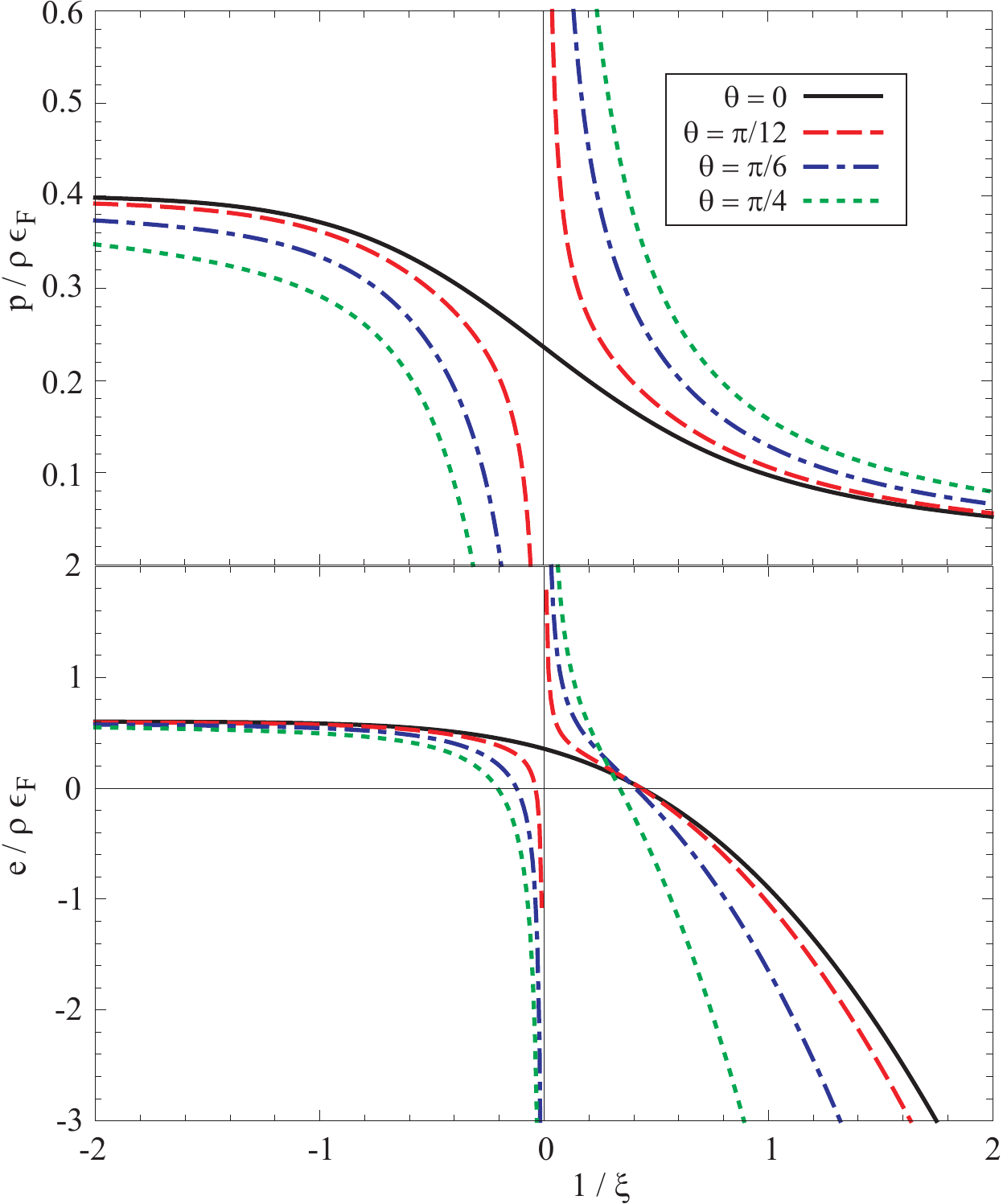}
   \caption{\label{f:T0pu-eta}(Color online) Zero temperature pressure and energy per unit volume
   \vs\ $1/\xi$ for several values of the parameter $\theta = 0$.} 
\end{figure}

\begin{figure}[t]
   \includegraphics[width=0.95\columnwidth]{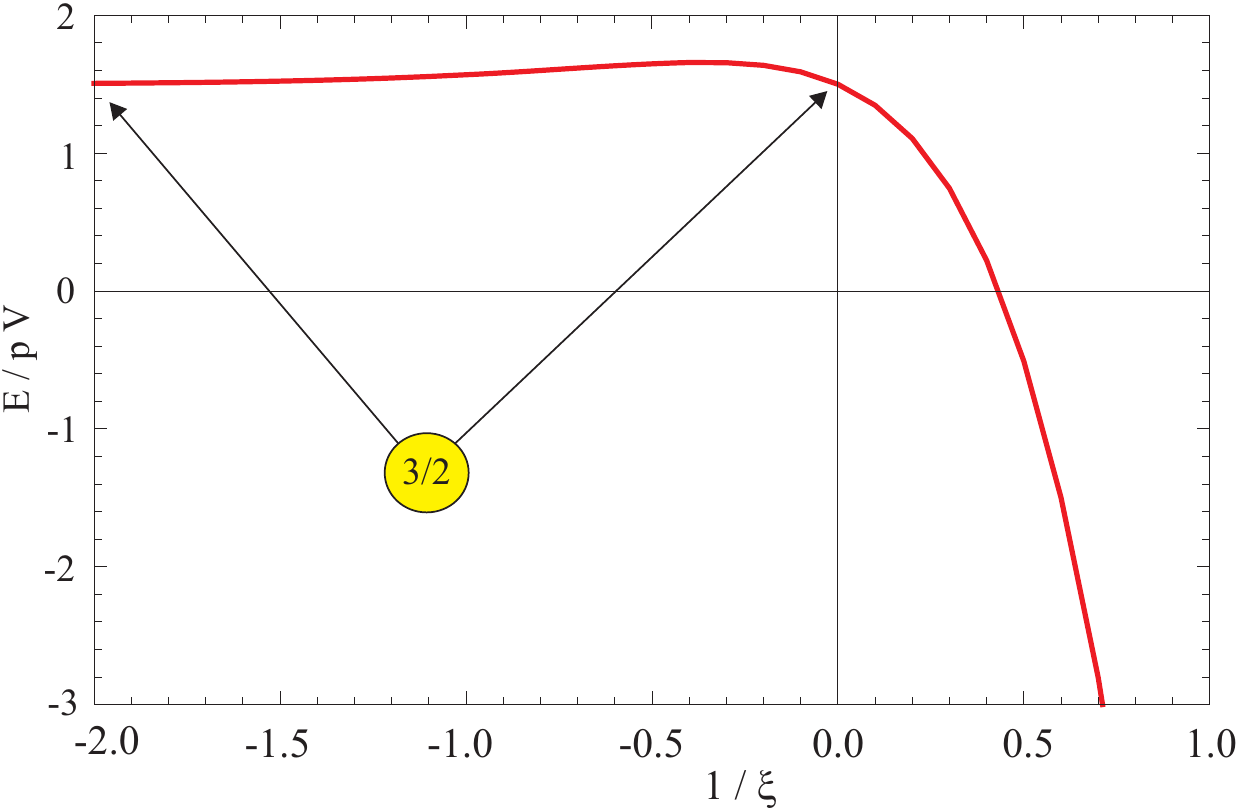}
   \caption{\label{f:T0theta0-ratio}(Color online) Zero temperature equation of state for $\theta = 0$ \vs\ $1/\xi$. We note that the ratio $e/p$ is equal to 3/2 
   both in the limit of a noninteracting Fermi gas and in the unitarity limit.}
\end{figure}

For illustrative purposes in Fig.~\ref{f:T0pu-eta} we depict the zero-temperature pressure and energy per unit volume as a function of $1/\xi$, for several values of~$\theta$.
The pressure and density have singularities in the unitarity limit, consistent with our previous results that the case of $\theta=0$ corresponds to the only physical theory for a dilute gas of fermions in the auxiliary field formalism.
In Fig.~\ref{f:T0theta0-ratio}, we illustrate the equation of state, $E / p V$, \vs\ $1/\xi$ for $\theta=0$.

%
%%%%%%%%%%%%%%%%%%%%%%%%%%%%%%%%%%%%%%%%%%%%%%%%%%%%%%%%%%%%%%%%%%%%%%
%
% Fig 3
%
%
%%%%%%%%%%%%%%%%%%%%%%%%%%%%%%%%%%%%%%%%%%%%%%%%%%%%%%%%%%%%%%%%%%%%%%
%
% Fig 4
%

%
%%%%%%%%%%%%%%%%%%%%%%%%%%%%%%%%%%%%%%%%%%%%%%%%%%%%%%%%%%%%%%%%%%%%%%
%

\subsection{\label{s:contact}Contact interaction relations}

For fermions interacting via short-range potential, Tan derived a set of universal relations in Refs.~\onlinecite{r:Tan:2008uq,*r:Tan:2008kx,*r:Tan:2008vn} that are independent of the details of the short-range interactions, some of which have been verified in experiments \cite{r:Stewart:2008zr,r:Stewart:2010ys}.  In particular, Tan relates the fermion momentum distribution, $\rho(k)$, at asymptotically large momentum to thermodynamics quantities such as the energy of the system per unit volume:
Tan showed~\cite{r:Tan:2008kx} that the fermion momentum distribution satisfies the property that 
\begin{equation}\label{tan1}
    \rho(k) \, \rightarrow \, \frac{C}{k^4}
    \>,
\end{equation}
in the large momentum limit, where $C$ is the contact density. This results was observed experimentally by Stewart \emph{et al.}~\cite{r:Stewart:2008zr}. Next, according to Tan's ``adiabatic sweep'' theorem~\cite{r:Tan:2008vn}, the variation of the energy per unit volume, $e$, with respect to the inverse scattering length is given by
\begin{equation}\label{tan2}
    \frac{d e}{d a^{-1}} \, = \,
    - \, \frac{\gamma}{2\pi} \, C
    \>.
\end{equation}
This result was also verified experimentally by Stewart \emph{et al.}~\cite{r:Stewart:2010ys}.

We will show here that the LOAF approximation satisfies these two Tan relations:
First, from Eq.~\eqref{XXX.e:rhokdef}, we find that indeed
\begin{equation}
    \rho(k) \, = \, \frac{C_\mathrm{LOAF}}{k^4} \, + \, \calO \Bigl ( \frac{1}{k^6} \Bigr )
    \>,
\end{equation}
with the LOAF contact density
\begin{equation}
    C_\mathrm{LOAF} \, = \,
    \frac{\Delta'^2}{4 \gamma^2} 
    \>.
\end{equation}
Second, we take the derivative of the energy per unit volume, $e$, given in Eq.~\eqref{XXX.e:u} with respect to the inverse scattering length. 
Using Eq.~\eqref{XXX.e:u} and recalling that at the minimum we have $\partial\Veff/\partial\chi^{\prime}=0$, $\partial\Veff/\partial\Delta'=0$, and $\partial\mathcal{V}_{eff}/\partial\mu=-\rho$, we find that
\begin{align}
    \frac{d e}{d a^{-1}} \, = \, &
    - \, \frac{\partial p}{\partial a^{-1}} 
    \\ \notag \, = \, &
    - \, \frac{\Delta'^2}{8\pi \, \gamma} \, = \, 
    - \, \frac{\gamma}{2\pi} \, C_\mathrm{LOAF}
    \>,
\end{align}
as indicated by Tan's relation~\eqref{tan2}.

%
%%%%%%%%%%%%%%%%%%%%%%%%%%%%%%%%%%%%%%%%%%%%%%%%%%%%%%%%%%%%%%%%%%%%%%
%
% Fig 5
%

%
%%%%%%%%%%%%%%%%%%%%%%%%%%%%%%%%%%%%%%%%%%%%%%%%%%%%%%%%%%%%%%%%%%%%%%
%

\subsection{\label{s:unitarity}Unitarity limit}

From Eqs.~\eqref{cIIA.e:gapdenI}, we see that in the unitarity limit, $1/\xi \rightarrow 0$,  the gap equations can only be satisfied if $\theta = 0$, which shows yet again that the case of $\theta=0$ corresponds to the only physical theory for a dilute gas of fermions in the auxiliary field formalism.
Hence, in the unitarity limit, the scaled gap equations become
\begin{subequations}\label{u.e:gapdenI}
\begin{gather}
   0
   =
   \frac{2}{\pi}
   \int_{0}^{\infty} \!\! \rd k \,
   \Bigl \{ \,
      1
      -
      \frac{ k^2}{ \omega_k } \,
      [ \, 1 - 2 \, n( \beta \omega_k ) \, ]
   \Bigr \} \>,
   \label{u.e:gapA} \\
   1
   =
   \frac{3}{2}
   \int_{0}^{\infty} \!\! k^2 \, \rd k \,
   \Bigl \{ \,
      1
      -
      \frac{ k^2 - \mu }{ \omega_k } \,
      [ \, 1 - 2 \, n( \beta \omega_k ) \, ]
   \Bigr \} \>,
   \label{u.e:gapB}
\end{gather} 
\end{subequations} 
where now $\omega_k^2 = ( k^2 - \mu )^2 + |\Delta|^2$, and we have dropped the bar notation.
The pressure and energy per unit colume are now given by
\begin{subequations}\label{u.e:pu}
\begin{align}
   \frac{p}{\rho \, \epsilonF}
   &=
   \frac{3}{2}
   \int_{0}^{\infty} \!\! k^2 \, \rd k \,
   \Bigl \{ \,
      \omega_k
      -      
      k^2
      +
      \mu
      -
      \frac{ | \Delta |^2 }{ 2 \, k^2 } \,
      \label{u.e:pressI} \\
      & \qquad\qquad
      +
      \frac{2}{\beta} \,
         \Ln{ 1 + e^{ -\beta \omega_k } }  \,
   \Bigr \} \>,
   \notag \\ 
   \frac{e}{\rho \, \epsilonF}
   &=
   \mu
   -
   \frac{3}{2}
   \int_{0}^{\infty} \!\! k^2 \, \rd k \,
   \Bigl \{ \,
      \omega_k \,
      [ \, 1 - 2 n( \beta \omega_k ) \, ]
      \label{u.e:energyI} \\
      & \qquad\qquad
      -  
      k^2
      +
      \mu 
      -
      \frac{ | \Delta |^2 }{ 2 \, k^2 } \,
   \Bigr \} \>.
   \notag    
\end{align}
\end{subequations}
By parts integration, we have
\begin{align}
   &\int_{0}^{\infty} \!\! k^2 \, \rd k \,
   \frac{1}{\beta} \,
   \Ln{ 1 + e^{ -\beta \omega_k } } 
   \label{u.e:bypartsI} \\
   & \qquad\qquad\qquad
   =
   \frac{2}{3}
   \int_{0}^{\infty} \!\! k^2 \, \rd k \,
   \frac{ k^2 \, ( k^2 - \mu ) }{ \omega_k } \, n( \beta \omega_k ) \>.
   \notag
\end{align}
Substituting this into Eq.~\eqref{u.e:pressI}, the pressure can be written as
\begin{align}
   \frac{p}{\rho \, \epsilonF}
   &=
   \int_{0}^{\infty} \!\! k^2 \, \rd k \,
   \Bigl \{ \,
      \frac{3}{2} \,
      \Bigl [ \,
         \omega_k
         -      
         k^2
         +
         \mu
         -
         \frac{ | \Delta |^2 }{ 2 \, k^2 } \,
      \Bigr ]
      \label{u.e:pressII} \\
      & \qquad\qquad
      +
      2 \,
      \frac{ k^2 \, ( k^2 - \mu ) }{ \omega_k } \, n( \beta \omega_k ) \,
   \Bigr \} \>.
   \notag    
\end{align}
For the energy expression, multiply Eq.~\eqref{u.e:gapB} by $\mu$ and substitute the result into Eq.~\eqref{u.e:energyI}.  This gives for the energy
\begin{align}
   \frac{e}{\rho \, \epsilonF}
   &=
   \frac{3}{2}
   \int_{0}^{\infty} \!\! k^2 \, \rd k \,
   \Bigl \{ \,
      -
      \frac{ \omega_k^2 + \mu ( k^2 - \mu) }{ \omega_k } \,
      [ \, 1 - 2 n( \beta \omega_k ) \, ]
      \notag \\
      & \qquad\qquad
      +      
      k^2
      +
      \frac{ | \Delta |^2 }{ 2 \, k^2 } \,
   \Bigr \} \>.
   \label{u.e:energyII}      
\end{align}
Now form the quantity
\begin{align}
   &\frac{ 2 e - 3 p }{ \rho \, \epsilonF }
   =
   3
   \int_{0}^{\infty} \!\! k^2 \, \rd k \,
   \Bigl \{ \,
      -
      \frac{ \omega_k^2 + \mu ( k^2 - \mu) }{ \omega_k }
     +      
      k^2
      +
      \frac{ | \Delta |^2 }{ 2 \, k^2 }
      \notag \\
      &
      -
      \frac{3}{2} \,
      \Bigl [ \,
         \omega_k
         -      
         k^2
         +
         \mu
         -
         \frac{ | \Delta |^2 }{ 2 \, k^2 } \,
      \Bigr ]
      +
      \frac{ 2 | \Delta |^2 }{ \omega_k } \, n( \beta \omega_k ) \,
   \Bigr \} \>.
   \label{u.e:calcI}      
\end{align}
Now integrate by parts,
\begin{align}
   &
   \int_{0}^{\infty} \!\! k^2 \, \rd k \,
   \Bigl [ \,
      \omega_k
      -      
      k^2
      +
      \mu
      -
      \frac{ | \Delta |^2 }{ 2 \, k^2 } \,
   \Bigr ]
   \label{u.e:bypartsII} \\
   &=
   \frac{2}{3}
   \int_{0}^{\infty} \!\! k^2 \, \rd k \,
   \Bigl [ \,
      k^2
      -
      \frac{ k^2 ( k^2 - \mu )}{\omega_k}
      -
      \frac{ | \Delta |^2 }{ 2 \, k^2 } \,
   \Bigr ] \>.
   \notag
\end{align}
Inserting this into \eqref{u.e:calcI} gives
\begin{equation}\label{u.e:calcII}
   \frac{ 2 e - 3 p }{ \rho \, \epsilonF }
   =
   3 \, | \Delta |^2
   \int_{0}^{\infty} \!\! \rd k \,
   \Bigl \{ \,
      1
      -
      \frac{ k^2 [ 1 - 2 \, n( \beta \omega_k ) ] }
           { \omega_k } \,
   \Bigr \} 
   =
   0 \>,
\end{equation}
where we have used the gap equation \eqref{u.e:gapA}.  So this shows that at the unitary limit,
\begin{equation}\label{u.e:purelation}
   e
   =
   \frac{3}{2} \, p \>,
\end{equation}
for all temperatures $T$ (see e.g. Ref.~\onlinecite{r:He:2007vn}). In Fig.~\ref{f:T0theta0-ratio} we show numerically that this relation holds for $T=0$.  

Using Eqs.~\eqref{XXX.e:Tzero-u-p} and \eqref{u.e:purelation}, we find the unitarity limit results at zero temperature, 
\begin{equation}\label{u.e:energyTzeroUnitarity}
   \frac{e}{\rho \, \epsilonF}
   =
   \frac{3}{5} \, \bar \mu \>,
\end{equation}
in reduced units.  But, at $T=0$, we have $\mu / \epsilonF = 0.59$. Therefore, introducing the energy per particle
\begin{equation}
    \varepsilon = \frac{E}{N} = \frac{e}{\rho}
    \>,
\end{equation}
we obtain that at zero temperature, in the unitarity limit,  we have
\begin{equation}
     (\varepsilon / \varepsilon_0)_\textrm{LOAF} = 0.59 
     \>.
\end{equation}

%
%%%%%%%%%%%%%%%%%%%%%%%%%%%%%%%%%%%%%%%%%%%%%%%%%%%%%%%%%%%%%%%%%%%%%%
%

%
%%%%%%%%%%%%%%%%%%%%%%%%%%%%%%%%%%%%%%%%%%%%%%%%%%%%%%%%%%%%%%%%%%%%%%
%
\section{\label{s:concl}Conclusions}

To summarize, in this paper we derived the auxiliary field formalism for a dilute fermionic atom gas with tunable interactions. This formalism represents the fermionic counterpart of a similar auxiliary field formalism introduced recently to describe the properties of a dilute gas of Bose particles~\cite{r:Cooper:2010fk}. Here we demonstrate that at zero temperature, the fermionic  LOAF equations are the same as the equations derived by Leggett~\cite{r:Leggett:1980fk}, whereas the finite-temperature results correspond to those discussed earlier by S\'a de~Melo, Randeria, and Engelbrecht~\cite{r:Melo:1993vn,r:Engelbrecht:1997fk}. The LOAF formalism shows that the BCS ansatz represents the \emph{only} physical auxiliary field theory for a dilute Fermi gas.  Furthermore, we showed that LOAF satisfies Tan's relation regarding the momentum distribution of fermions  at asymptotically large momenta and Tan's ``adiabatic sweep'' theorem. Just like in the Bose case, the auxiliary field approach for fermions provides a systematic framework that allows one to improve the LOAF results by  calculating the 1-PI action corrections, order by order. 

\bigskip
\begin{acknowledgments}
This work was performed in part under the auspices of the U.~S.~Dept.~of Energy.  The authors would like to thank the Santa Fe Institute for hospitality during this work.  JFD would like to thank LANL for travel support and hospitality.
\end{acknowledgments}

%
%%%%%%%%%%%%%%%%%%%%%%%%%%%%%%%%%%%%%%%%%%%%%%%%%%%%%%%%%%%%%%%%%%%%%%%
%
\bibliography{johns}
%
%%%%%%%%%%%%%%%%%%%%%%%%%%%%%%%%%%%%%%%%%%%%%%%%%%%%%%%%%%%%%%%%%%%%%%
%
\end{document}